\documentclass[runningheads]{llncs}

\usepackage{cite}
\usepackage{amsmath,amssymb,amsfonts}
\usepackage{algorithmic}
\usepackage{graphicx}
\usepackage{textcomp}

\usepackage[english]{babel}
\usepackage{xurl}
\usepackage{listings}
\usepackage[inline,shortlabels]{enumitem}
\usepackage{adjustbox}
\usepackage{carshapes}
\usepackage{csquotes}
\usepackage{booktabs}
\usepackage{makecell}
\usepackage{todonotes}
\usepackage[draft]{tikzpeople}
\usepackage[nohyperlinks, printonlyused]{acronym}
\usepackage{multirow}
\usepackage{pgfplots}
\pgfplotsset{compat=1.3}
\usepackage[hidelinks]{hyperref}
\usepackage[capitalise,nameinlink]{cleveref}
\Crefname{section}{Sec.}{Secs.}
\Crefname{figure}{Fig.}{Figs.}
\Crefname{table}{Tab.}{Tabs.}
\Crefname{definition}{Def.}{Defs.}
\usepackage{tikz}
\usetikzlibrary{positioning, shapes, arrows, fit, trees, matrix, chains, calc}
\tikzstyle{comm} = [thick,<->,>=stealth,align=center]
\tikzstyle{arrow} = [thick,->,>=stealth,align=center]
\tikzstyle{pki} = [semithick,rectangle, text width=5.5em, minimum width=6.5em, minimum height=4em, text centered, draw=black]
\tikzstyle{opt} = [thick,->,>=stealth,dashed]
\tikzstyle{box} = [draw, dotted]
\tikzstyle{actor} = [semithick,rectangle, minimum width=3em, minimum height=4.5em, text centered, draw=black,align=center, text width=6em,fill=white]
\tikzstyle{backend} = [cloud, cloud puffs=17.5, cloud ignores aspect, semithick, minimum width=10em, minimum height=3.2em, text centered, draw=black,align=center,fill=white, inner xsep=0.05em, inner ysep=0.05em]
\newcommand{\subscript}[2]{$#1 _ #2$}

\newcommand{\rfc}[1]{RFC #1 \cite{RFC#1}}

\newcommand\fig[1]{Figure~\ref{fig:#1}}

\newcommand\sect[1]{Section~\ref{sec:#1}}

\begin{document}

\title{Streamlining Plug-and-Charge Authorization for Electric Vehicles with OAuth2 and OIDC}

%
%
%

\titlerunning{Streamlining PnC EV Charging with OAuth2 \& OIDC}
    \author{%
        Jonas Primbs\inst{1}\orcidID{0000-0001-5479-3315}%
    \and
        Dustin Kern\inst{2}\orcidID{0000-0003-4365-1762}%
    \and
        Michael Menth\inst{1}\orcidID{0000-0002-3216-1015}%
    \and
        Christoph Krauß\inst{2}\orcidID{0000-0001-7776-7574}%
    }
    \authorrunning{J. Primbs et al.}
    \institute{%
        Eberhard Karls Universität, Tübingen, Germany %
            \email{\{jonas.primbs,michael.menth\}@uni-tuebingen.de}%
    \and%
        Darmstadt University of Applied Sciences, Darmstadt, Germany %
            \email{\{dustin.kern,christoph.krauss\}@h-da.de}%
    }
\maketitle

\begin{abstract}
    The \ac{PnC} process defined by ISO~15118 standardizes automated \ac{EV} charging by enabling automatic installation of credentials and use for authentication between \ac{EV} and \ac{CP}. 
However, the current credential installation process is non-uniform, relies on a complex \ac{PKI}, lacks support for fine-grained authorization parameters, and is not very user-friendly. 
In this paper, we propose a streamlined approach to the initial charging authorization process by leveraging the OAuth Device Authorization Grant and Rich Authorization Requests.
The proposed solution reduces technical complexity, simplifies credential installation, introduces flexible authorization constraints (e.g., time- and cost-based), and facilitates payment through \ac{OIDC}.
We present a proof-of-concept implementation along with performance evaluations and conduct a symbolic protocol verification using the Tamarin prover.
Furthermore, our approach solves the issue of OAuth's cross-device authorization, making it suitable as a formally proven blueprint in contexts beyond \ac{EV} charging.

\end{abstract}


%
\acresetall
\section{Introduction}
\label{sec:introduction}
The charging of \acp{EV} at public \acp{CP} is currently realized with different systems for user authorization, e.g., RFID cards or charging apps.
To make this process more user-friendly, various protocols have been defined in recent years.
One of the most important developments is the definition of \ac{PnC} based on the ISO~15118 standard \cite{iso2}.
With \ac{PnC}, an \ac{EV} can use locally installed credentials, i.e., a certificate and corresponding key pair, to authenticate itself at a \ac{CP} without requiring user interaction.
This \ac{PnC} authentication then serves as the basis for charge authorization and billing.

The \ac{PnC} process, however, requires a complex \ac{PKI} where several backend actors operate individual (sub-) \acp{CA}.
The custom ISO~15118 \ac{PKI} and corresponding credential installation processes impose a high level of complexity, which is reflected in the critique of relevant stakeholders \cite{isoPkiConsiderations} and which means that complexity-reducing measures are important \cite{ELAADNL2018}. 
Moreover, relevant parts of the installation process are out-of-scope for ISO~15118.
This resulted in standardization bodies trying to fill the gaps \cite{vde-ar-2802} and in a non-uniform \ac{PnC} installation process which varies based on the combination of \ac{EV} model and \ac{EMSP}.
Usability for end users should also be simplified when installing contract credentials.

In this paper, we address these issues by proposing a novel method based on OAuth 2 that streamlines the credential installation process for \ac{EV} charging.
Compared to the patchwork of proprietary solutions that are utilized for the current credential management method, OAuth 2 represents a widely accepted and standardized solution.
The proposed approach reduces the complexity of the required infrastructure by reducing the number of required sub-\acp{CA} in the \ac{PKI}.
It simplifies the installation process for drivers with a uniform app-guided process which is independent of \ac{EV} model and \ac{EMSP} and utilizes OAuth's \acf{RAR} to increase security by enabling more fine-grained authorizations like limiting the \ac{EV}'s charging periods/costs.
The proposed method can be extended by an \ac{OIDC}-based login with payment providers, as known from the express checkout feature in online stores.
This extension enables drivers to become a customer of the \ac{EMSP} without creating a new account.
The main contributions of the paper can be summarized as follows.
\begin{enumerate*}[label=\emph{(\roman*)}]
	\item We analyze the current \ac{EV} charging processes, obtaining a relevant adversary model, as well as security and feasibility requirements with respect to an OAuth-based solution.
	\item We design a detailed concept for the application of OAuth methods to \ac{EV} charging under consideration of the respective adversary and requirements.
	\item We implement the proposed solution as a proof-of-concept, evaluate it with regard to feasibility, and publish the corresponding source code.
	\item We verify the security of the proposed solution using a tool-based formal security verification with the Tamarin prover \cite{meier2013tamarin} and publish the corresponding Tamarin models.
\end{enumerate*}

The remainder of the paper is structured as follows.
In \cref{sec:background}, we provide relevant background information for the \ac{EV} charging architecture and for OAuth~2.
Afterwards, we discuss related work in \cref{sec:rel}.
The adversary model and identified requirements are described in \cref{sec:model}.
In \cref{sec:concept}, we introduce the concept of OAuth-based \ac{EV} charging authorization and the \ac{OIDC} extension for online payment.
Then, we provide information about the proof-of-concept implementation in \cref{sec:impl} and the performance and security evaluation in \cref{sec:eval}.
We conclude the paper in \cref{sec:conclusion}. 

\section{Background}
\label{sec:background}
In this section, we describe the necessary background on \ac{EV} charging and the OAuth 2 authorization framework to understand the paper.

\subsection{\acs*{EV} Charging Architecture}
\label{sec:emob}

\fig{systemmodel} shows a simplified \ac{EV} charging architecture based on the ISO~15118 charge communication standard~\cite{iso2}.
Besides the \acf{EV} and \acf{CP}, the system for contract-based charging involves:
\begin{enumerate*}[label=\emph{(\roman*)}]
	\item the \acf{OEM} which produces the \ac{EV},
	\item the \acf{CPO} which manages \acp{CP} for billing and load management,
    \item the \acf{CPS} which acts as a trusted third-party during credential installation, and
	\item the \acf{EMSP} with which an \ac{EV} user has a contract for charge authorization and billing.
\end{enumerate*}

\pgfdeclareimage[width=1.5em]{cp}{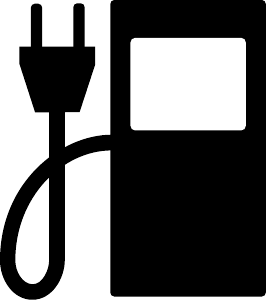}
\begin{figure}[h]
	\centering
				\begin{tikzpicture}
						\node (EV) [actor, text width=3.5em, minimum height=6em] {\acf*{EV}};	
						\node (CP) [actor, right=8.5em of EV, text width=3.5em, minimum height=6em] {\acf*{CP}};	
						\node (CPO) [actor,below=7.8em of CP, text width=6em,xshift=-3.5em] {\acf*{CPO}};
						\node (CPS) [actor,left=8.5em of CPO, text width=5.5em] {\acf*{CPS}};
						\node (EMSP) [actor,left=0.0em of CPS, text width=6.0em] {\acf*{EMSP}};
						\node (OEM) [actor,left=6.5em of EV, text width=6.5em] {\acf*{OEM}};
						\node (space) [right=2.7em of CP] {};
						
						\node [minimum width=2.5em,yshift=3.0em,xshift=1.1em] at (CP) {\includegraphics[width=1.5em]{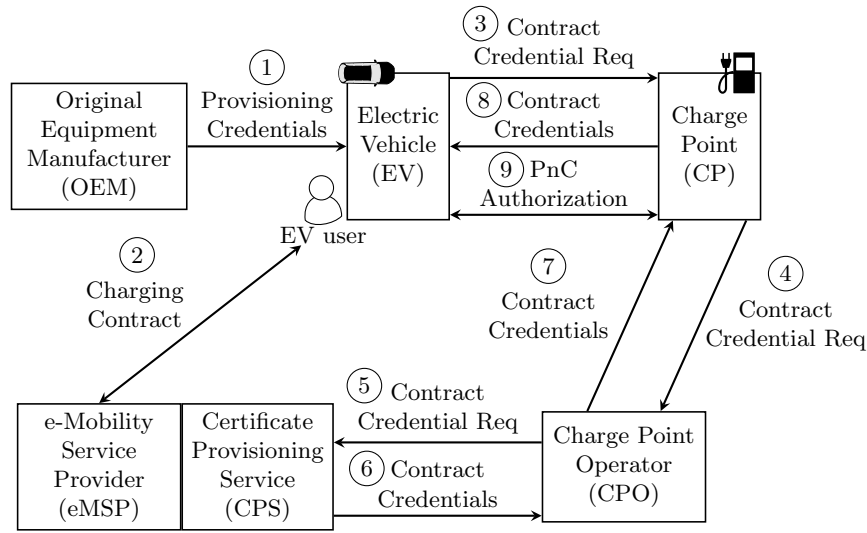}};
						\node [sedan top,draw=black,body color=black!,minimum width=2.5em,yshift=3.0em,xshift=-1.3em] at (EV) {}; 
						
						
						\node (EVu) [duck, left=.2em of EV,yshift=-2.2em,minimum size=1.5em, align=center] {\small \acs*{EV} user};
						
						
						\draw[arrow](OEM) -- node (m1) [above] {Provisioning \\ Credentials} (EV);
						
						\draw[arrow,transform canvas={yshift=2.8em}](EV) -- node (m2) [above] {Contract \\ Credential Req} (CP);

						\draw[arrow,transform canvas={xshift=1.5em}](CP.270) -- node (m3) [right,transform canvas={xshift=-0.3em,yshift=-0.4em}] {Contract \\ Credential Req} (CPO.90);						
						\draw[arrow,transform canvas={xshift=-1.5em}](CPO.90) -- node (m4) [left,transform canvas={xshift=-0.5em,yshift=-0.0em}] {Contract \\ Credentials} (CP.270);
						\draw[arrow,transform canvas={yshift=1.0em}](CPO) -- node (m31) [above] {Contract \\ Credential Req} (CPS);						
						\draw[arrow,transform canvas={yshift=-2.0em}](CPS) -- node (m41) [above] {Contract \\ Credentials} (CPO);

						\draw[arrow,transform canvas={yshift=-0.0em}](CP) -- node (m5) [above] {Contract \\ Credentials} (EV);
						\draw[comm,transform canvas={yshift=-2.8em}](CP) -- node (m6) [above] {\acs*{PnC} \\ Authorization} (EV);
						
						\node (char) [shape=circle,draw,inner sep=2pt, yshift=0.0em, above=-0.1em of m1] {1};
						
						\node (char) [shape=circle,draw,inner sep=2pt, yshift=-0.0em, xshift=-3.0em, above=1.5em of m2] {3};
						\node (char) [shape=circle,draw,inner sep=2pt, yshift=-0.4em, xshift=1.2em, above=0.0em of m3] {4};
						\node (char) [shape=circle,draw,inner sep=2pt, yshift=0.0em, xshift=-2.0em, above=0.0em of m4] {7};
						\node (char) [shape=circle,draw,inner sep=2pt, yshift=-0.5em, xshift=-3.0em, above=0.3em of m31] {5};
						\node (char) [shape=circle,draw,inner sep=2pt, yshift=-0.5em, xshift=-2.9em, above=-2.9em of m41] {6};
						
						\node (char) [shape=circle,draw,inner sep=2pt, yshift=0.6em, xshift=1.3em, left=0.6em of m5] {8};
						\node (char) [shape=circle,draw,inner sep=2pt, yshift=-2.25em, xshift=1.3em, left=-0.9em of m6] {9};

						\draw[arrow, <->,transform canvas={xshift=-0.0em}]([xshift=-0.9em,yshift=-0.6em]EVu.270) -- node (m00) [left,transform canvas={xshift=-0.3em,yshift=0.8em}] {Charging\\Contract} (EMSP.90);	
                        \node (char) [shape=circle,draw,inner sep=2pt, yshift=0.8em, xshift=-0.2em, above=0.0em of m00] {2};


				\end{tikzpicture}
	\caption{\acs*{EV} charging architecture.} 
	\label{fig:systemmodel}
\end{figure}

The architecture in \fig{systemmodel} allows an \ac{EV} user to receive energy from different suppliers at a \ac{CP}.
The supplier in this case is the \ac{EMSP} with which the user has concluded a contract before charging.
The \ac{CPO} then invoices the \ac{EMSP} for the consumed energy, and the \ac{EMSP} invoices the \ac{EV} user.

The conclusion of the contract between the user and the \ac{EMSP} is not standardized and therefore varies by provider. For example, registration might occur through a website or app. Our work proposes a procedure for this task in such a way that the contract can also be concluded just before charging and, if desired, also limited to a maximum payment amount or a maximum time (more details in \cref{sec:concept}).

Each \ac{EV} has a certificate and key pair to authenticate itself. 
This is referred to as provisioning credentials and is issued by the \ac{EV}'s \ac{OEM}.
To charge at a \ac{CP}, the \ac{EV} needs a contract credential (another certificate and key pair) from the \ac{EMSP}, proving that the user has a valid charging contract.

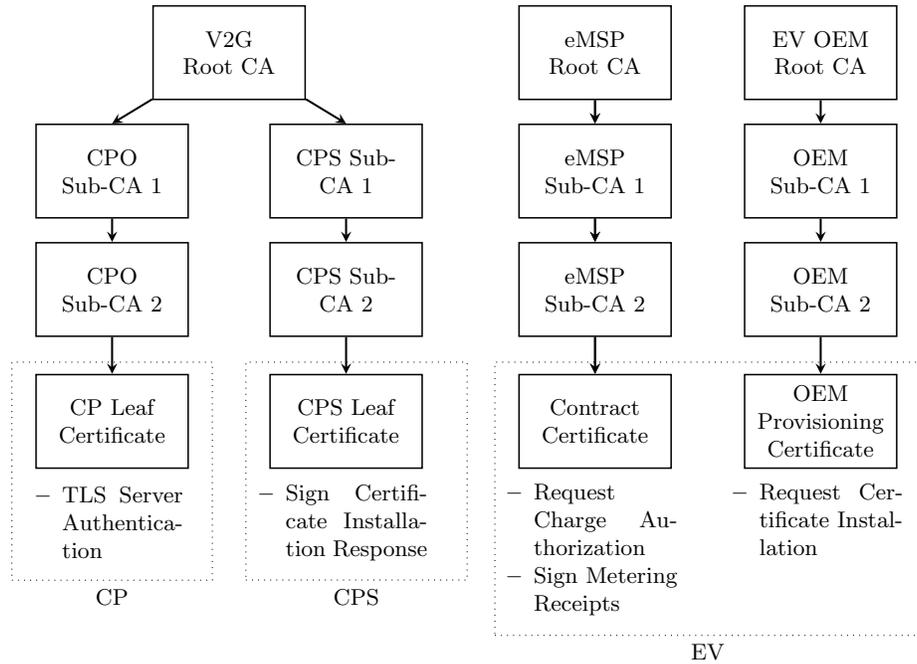
\begin{figure}[h]
	\centering
	\adjustbox{max width=1\linewidth}{
		{\small \begin{tikzpicture}
				\node (R) [pki] {V2G Root CA};
				\node (moR) [pki, right= of R, xshift=6em] {\acs*{EMSP} Root CA};
				\node (oemR) [pki, right= of moR, xshift=0em] {EV OEM Root CA};
				
				\node (cpo1) [pki, below= 1em of R, xshift=-5em] {\acs*{CPO} Sub-CA 1};
				\node (cpo2) [pki, below= 1em of cpo1] {\acs*{CPO} Sub-CA 2};
				\node (evse) [pki, below= of cpo2, yshift=1.5em] {\acs*{CP} Leaf Certificate};
				
				\node (prov1) [pki, below= 1em of R, xshift=5em] {\acs*{CPS} Sub-CA 1};
				\node (prov2) [pki, below= 1em of prov1] {\acs*{CPS} Sub-CA 2};
				\node (prov) [pki, below= of prov2, yshift=1.5em] {\acs*{CPS} Leaf Certificate};
				
				\node (mo1) [pki, below= 1em of moR] {\acs*{EMSP} Sub-CA 1};
				\node (mo2) [pki, below= 1em of mo1] {\acs*{EMSP} Sub-CA 2};
				\node (cc) [pki, below= of mo2, yshift=1.5em] {Contract Certificate};
				
				\node (oem1) [pki, below= 1em of oemR] {OEM Sub-CA 1};
				\node (oem2) [pki, below= 1em of oem1] {OEM Sub-CA 2};
				\node (pc) [pki, below= of oem2, yshift=1.5em] {OEM Provisioning Certificate};
				
				\node (evseL) [below= -1.4em of evse,yshift=-1.7em, xshift=-0.3em] {\makebox(3em,3.25em){\small
						{\begin{varwidth}{8em}\begin{itemize}
									\item TLS Server Authentication
						\end{itemize}\end{varwidth}}
				}};
				
				\node (provL) [below= -1.4em of prov,yshift=-1.7em, xshift=-0.3em] {\mbox{\small
						{\begin{varwidth}{7.9em}\begin{itemize}
									\item Sign Certificate Installation Response
						\end{itemize}\end{varwidth}}
				}};
				
				\node (ccL) [below= -1.4em of cc,yshift=-1.7em, xshift=-0.3em] {\mbox{\small
						{\begin{varwidth}{8em}\begin{itemize}
									\item Request Charge Authorization
									\item Sign Metering Receipts
						\end{itemize}\end{varwidth}}
				}};
				
				\node (pcL) [below= -1.4em of pc,yshift=-1.7em, xshift=-0.3em] {\mbox{\small
						{\begin{varwidth}{8em}\begin{itemize}
									\item Request Certificate Installation
						\end{itemize}\end{varwidth}}
				}};
				
				\node[box,inner xsep=1em,inner ysep=0.5em,label=below:CP,fit=(evse.north west) (evse.north east) (evseL.south east)] {};
				\node[box,inner xsep=1em,inner ysep=0.5em,label=below:CPS,fit=(prov.north west) (provL.south east)] {};
				\node[box,inner xsep=1em,inner ysep=0.5em,label=below:EV,fit=(cc.north west) (ccL.south east) (pc.north east) (pcL.south west)] {};
				
				\draw[arrow](R) -- (cpo1.north);
				\draw[arrow](R) -- (prov1.north);
				
				\draw[arrow](cpo1) -- (cpo2);
				\draw[arrow](cpo2) -- (evse);
				
				\draw[arrow](prov1) -- (prov2);
				\draw[arrow](prov2) -- (prov);
				
				\draw[arrow](moR) -- (mo1);
				\draw[arrow](mo1) -- (mo2);
				\draw[arrow](mo2) -- (cc);
				
				\draw[arrow](moR) -- (mo1);
				\draw[arrow](mo1) -- (mo2);
				\draw[arrow](mo2) -- (cc);
				
				\draw[arrow](oemR) -- (oem1);
				\draw[arrow](oem1) -- (oem2);
				\draw[arrow](oem2) -- (pc);
	\end{tikzpicture}}}
	\caption{ISO~15118 \acs*{PKI} (cf.~\cite{iso2}).}
	\label{fig:isoPKI}
\end{figure}

To enable the installation of contract credentials into an \ac{EV}, the user provides the \ac{EV}'s unique provisioning credential ID to the \ac{EMSP} during the conclusion of the charging contract.
Afterwards, the \ac{EV} can generate and send a contract credential request to the \ac{EMSP}.
This contract credential request contains the \ac{EV}'s provisioning credential ID, which allows the \ac{EMSP} to identify the associated charging contract. 
The \ac{EMSP} then generates and returns the desired contract credentials to the \ac{EV}. 
With these credentials, the \ac{EV} can authenticate itself at \acp{CP} and be charged.
In the following, we explain these processes in detail: 
\begin{description}
	\item[\acs*{EV} Provisioning.] 
	During \ac{EV} production, the \ac{OEM} generated provisioning credentials for the \ac{EV} (cf. \cref{fig:systemmodel}, Step~1).
    The provisioning credentials consist of a key pair and public key certificate. The certificate contains a provisioning credential ID, which uniquely identifies the \ac{EV}.
    The OEM issues these credentials using its own certificate chain with an \ac{OEM} root \ac{CA} and up to two \ac{OEM} sub-\acp{CA} (cf. \cref{fig:isoPKI}).\newline
	Once the \ac{EV} is purchased, the \ac{EV} user can retrieve the provisioning credential ID from the \ac{EV}. A specific method for retrieving this ID is not defined.
    Using this ID, the user can register the \ac{EV}'s provisioning credentials with a charging contract at an \ac{EMSP} (cf. \cref{fig:systemmodel}, Step~2). This allows the \ac{EMSP} to create a link between the user/charging contract and the \ac{EV}/provisioning credentials, which is important for the contract credential installation process.

	\item[Contract Credential Installation.]
	When the \ac{EV} is connected to a \ac{CP}, communication starts over the charging cable using \ac{PLC}.
    The \ac{CP} is authenticated to the \ac{EV} using \ac{TLS} with the \ac{CP}'s leaf certificate.
	The \ac{CP}'s certificate is verified by the \ac{EV} through the \ac{CPO}'s certificate chain below the so-called \ac{V2G} root certificate (cf. \cref{fig:isoPKI}). The \ac{V2G} root certificate must be installed on the \ac{EV}. \newline
	After communication is set up, the \ac{EV} can send a contract credential installation request to the \ac{CP}.
    The \ac{CP} forwards the request over the \ac{CPO} to the \ac{EMSP} (cf. \cref{fig:systemmodel}, Steps~3--5).
    The contract credential installation request includes the \ac{EV}'s provisioning credential certificate/-ID and is signed with the \ac{EV}'s provisioning credential private key.
	The \ac{EMSP} verifies the signature based on the provisioning certificate, validates the provisioning certificate based on the \ac{OEM}'s root, and makes sure that the provisioning credential ID is registered with a user's charging contract. \newline 
    After the \ac{EMSP} has validated the \ac{EV}'s contract credential installation request, the \ac{EMSP} generates contract credentials for the \ac{EV}.
    That is, the \ac{EMSP} generates a key pair and issues the corresponding public key certificate.
    For issuing the contract credential certificate, the \ac{EMSP} has its own certificate chain with an \ac{EMSP} root \ac{CA} and up to two \ac{EMSP} sub-\acp{CA} (cf. \cref{fig:isoPKI}).
    Additionally, the private key of the contract credentials is encrypted for the \ac{EV} based on its provisioning credential public key. \newline 
	The \ac{EMSP} builds a contract credential installation response message that contains the contract credential certificate and the encrypted public key.
    This message is passed to the \ac{CPS} which signs the message for the \ac{EV}.
    For this, the \ac{CPS} possesses a \ac{CPS} leaf certificate, which is in a chain below a \ac{CPS} sub-\ac{CA} and below the \ac{V2G} root. \newline 
    Finally, the signed contract credential installation response message is passed over the \ac{CPO} and the \ac{CP} to the \ac{EV} (cf. \cref{fig:systemmodel}, Steps~6--8).
    After receiving the contract credential installation response message, the \ac{EV} verifies the \ac{CPS}' signature and decrypts the contract credential private key using its provisioning credential private key.
	The \ac{EV} then stores the encrypted contract credential private key and corresponding public key certificate for later uses.

	\item[\acs*{PnC} Authentication.]
	If the \ac{EV} is in possession of valid contract credentials, it can use these credentials for \ac{PnC} authentication at the \ac{CP} via a challenge-response protocol (cf. \cref{fig:systemmodel}, Step~9).
    For this purpose, the \ac{CP} sends a challenge (nonce) and the \ac{EV} responds by signing the nonce with its contract credential private key.
    The \ac{CP} verifies this signature with the public key from the \ac{EV}'s contract certificate whose validity is checked with the \ac{EMSP}'s certificate chain.
    This validity check requires the \ac{EMSP} root certificate to be installed on the \ac{CP}.
    If all these validations pass, the \ac{EV} is authorized to charge at the \ac{CP}.
    During the charging session, the \ac{EV} may sign the \ac{CP}'s energy meter readings with its contract credential private key, which serves as proof to the \ac{CPO} for billing with the \ac{EMSP}.
	The \ac{EMSP} uses the information received from the \ac{CPO} to bill the \ac{EV} user based on the existing charging contract.
\end{description}

\subsection{OAuth~2.0}
\label{sec:oauth2}
The OAuth~2 authorization framework is standardized in \rfc{6749}.
It is based on the \acf{HTTP} \cite{rfc7231} and the \acf{JSON} \cite{rfc8259}.
The framework enables users to grant their clients, which may be web or native applications, scoped access to \acfp{PR} on a \acf{RS}.
\acp{PR} may be user data or restricted actions; the user is therefore designated as the \acf{RO}.

Non-OAuth approaches often use a 'naive' approach for authentication and authorization, where the user stores the credentials in the client, thereby allowing the client direct access to all \acp{PR} at the \ac{RS}.
In contrast, the OAuth~2 protocol requires the user (\ac{RO}) to log in to an \acf{AS} and subsequently grant the client permission to access a pre-defined set of \acp{PR}, referred to as the scope.
Subsequently, the \ac{AS} issues an \acf{AT} to the client, which enables the client to access the \acp{PR} within the granted scope on the \ac{RS}.
OAuth thus enhances security by limiting client access to the user-granted scope within the user's account without disclosing credentials to the client.

\begin{description}
    \item[Device Authorization.] 
\label{sec:oauth_device_authorization}
\rfc{8628} extends OAuth 2 by a cross-device authorization flow in which the \ac{RO} authorizes a so-called device client, e.g., a streaming service on a smart TV, from a trusted device, e.g., a smartphone.
The flow's major advantage is that the \ac{RO} does not need to enter any credentials on the client device.
This improves usability because active sessions on the \ac{RO}'s device (the smartphone) can be re-used.
It also improves security because the device client only obtains a scoped access token instead of a full-fledged session identifier.
\fig{device_authorization} illustrates the typical cross-device authorization flow.

\begin{figure}[h]
    \centering
    \includegraphics[width=1\linewidth]{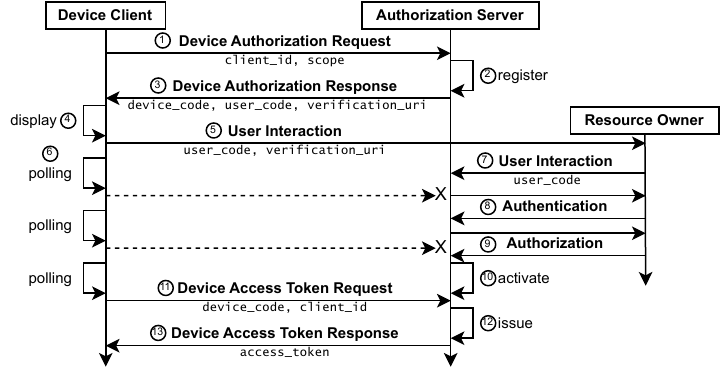}
    \caption{OAuth 2 cross-device authorization flow.}
    \label{fig:device_authorization}
\end{figure}

First, the device client sends a device authorization request to the \ac{AS}, containing its client ID which was pre-registered by a developer, and the requested scope (1).
The \ac{AS} then registers this request (2), responds with a verification URI, and issues a complex non-guessable device code, and a non-guessable but easily transcribable user code of typically 8 alphanumeric characters (3).
The device client then displays (4) the verification URI typically in form of a QR code and the device code to the \ac{RO} (5).
Then, it starts polling a device access token request from the \ac{AS} periodically (6).

The \ac{RO} then browses the verification URI on a trusted device and enters the displayed user code (7).
Subsequently, the \ac{AS} authenticates the \ac{RO} (8) who then reviews the requested scope to determine whether to grant authorization to the device client (9).
When granted, the \ac{AS} activates the corresponding device code (10).
At the next time, when the device client polls a device access token request with this device code and the corresponding client ID (11), the \ac{AS} generates an access token (12) and issues it in the device access token response (13).
This  token contains the authorized scope with a short-term validity period and is signed with \ac{AS}'s private key $K^-_{AS}$.

    \item[\acf{RAR}.] 
\label{sec:rar}
In OAuth~2, scopes are finite, meaning they can, e.g., grant payment authorization in general, but they cannot grant specific transaction limits.
This is because an infinite number of pre-registered scopes would be required to accommodate any given transaction amount.
\rfc{9396} introduces a \acf{RAR} which enables clients to incorporate supplementary authorization details in the authorization request (1).
In the authorization step (9), the \ac{AS} will display these details to the \ac{RO} and include them in the issued \ac{AT} (13).
Such authorization details may include, e.g., a transaction limit of \$100 per month.

    \item[Resource Request.] 
With an \ac{AT} of an appropriate scope and authorization details, a client can access \acp{PR} on a \ac{RS}.
\fig{oauth_resource_request} illustrates a resource request.
The client transmits a resource request for the \acp{PR} to the \ac{RS} (1) which then validates the \ac{AT} and determines whether it has a sufficient scope and authorization details (2).
If valid, the \ac{RS}  responds with the \acp{PR} (3).
\begin{figure}[h]
    \centering
    \includegraphics[width=1\linewidth]{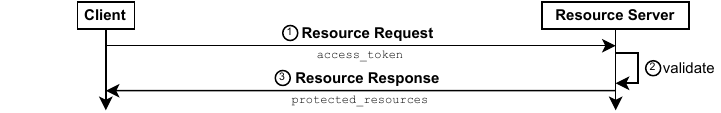}\vspace{-0.5em}
    \caption{OAuth 2 Resource Request.}
    \label{fig:oauth_resource_request}
\end{figure}
\end{description}

\section{Related Work}
\label{sec:rel}
In this section, we briefly summarize related work in the context of \ac{EV} charging security and OAuth 2.

\subsection{\acs*{EV} Charging Security}
\label{sec:rel:ev}
As the connectivity of modern vehicles increases, so does their potential attack surface \cite{plappert2021attack}.
One relatively new area of vehicle connectivity is that between an \ac{EV} and the charging infrastructure.
In this e-mobility system, there are several actors and communication protocols, which lead to various potential security threats \cite{antoun2020detailed}.
The possible consequences of attacks in the e-mobility sector range from financial damage over privacy violations to safety risks and even power grid stability issues \cite{falk2012electric,fuchs2020securing,kern2021analysis}.

Several papers propose security architectures for the \ac{EV} charging system that address different kinds of potential threats.
For example, \cite{fuchs2020a,fuchs2020hip,fuchs2020hip20} propose the integration of \acp{HSM} into ISO~15118 to protect the private billing-relevant data of \acp{EV} and ensure the integrity of an \ac{EV}'s software state.
The authors of \cite{van2015securing} focus on providing end-to-end security of charging data via middleware-based publish/subscribe communication, \cite{van2019non} looks at long-term non-repudiation, and \cite{kern2023quantumcharge} offers post-quantum security for \ac{EV} charging.
Various decentralized approaches exist:  \cite{Firoozjaei2019EVChain:Charging,Xu2021EVchain:Vehicles,knirsch2018privacy} propose different blockchain-based approaches to \ac{EV} charging  and
\cite{richter2021exploring,Hoess2022WithOW,kailus2024self} consider \acl*{SSI}-based methods.
Furthermore, several papers offer privacy enhancements in the \ac{EV} charging context. For instance, \cite{li2016portunes} proposes a pseudonym based fast authentication scheme for contactless charging, \cite{yucel2019efficient} designs a homomorphic encryption-based scheme for energy supplier matching, and the solutions presented in \cite{kern2022integrating,zelle2018anonymous,zhao2015secure} all look at the use of \acl*{DAA} schemes for \ac{EV} authentication.

In comparison to related work, this paper provides scientific novelty by proposing an OAuth-based approach to \ac{EV} credential handling, which offers secure fine-grained authorizations based on a widely used standardized framework (i.e., OAuth 2).
\subsection{OAuth 2}
\label{sec:rel:oauth}
OAuth was combined with EV charging several times.
Henry Gadacz proposed the Online Internet Account Authentication approach \cite{gadacz2021evaluation} where the driver authorizes payment via Google Pay with OAuth.
Therefore, the driver scans a QR code on the \ac{CP} which initializes an OAuth flow to authorize the \ac{CP} to debit money from the driver's bank account for the charging process.
This requires the \ac{CP} to implement this payment method, and the driver must trust the \ac{CP} and QR code to not being manipulated.
The driver must repeat this authorization process for every charging process.
In our approach, the driver authorizes the \ac{EV} for \ac{PnC} to debit money from the driver's bank account once and to pay for any future charging process.
Timpner et al. proposed a smartphone-based registration and key deployment for vehicle-to-cloud communications \cite{timpner2013secure} to persist a payment authorization for \ac{EV} charging processes.
In their proposal, the driver generates a key pair on a smartphone and requests a signed certificate from a \ac{CA} to deploy this key pair with the issued certificate to the \ac{EV}.
The EV can then authenticate with this certificate to prove authorization to pay for charging with the driver's bank account.
The approach is not directly compatible with the \ac{PnC} infrastructure, but some basic concepts may also be applied.
However, generating the key pair on a third-party device and then transferring it to the \ac{EV} is a bad design \cite{fuchs2020hip,fuchs2020hip20}.
In our approach, we fix this and implement modern OAuth extensions like the Device Authorization Grant (\rfc{8628}) and Rich Authorization Requests (\rfc{9396}) to improve the security further.
The Client-Initiated Backchannel Authentication (CIBA) Flow \cite{ciba} extends OpenID Connect by the capability of cross-device authentication and is typically used for online banking applications.
Applied to the \ac{EV} charging authorization use case, this would require the user to already have established a side channel from his smartphone to the \ac{EMSP} to receive push notifications.
It will also result in the user logging in with a cross-device flow on the \ac{EV} and authorizing the requested scopes there.
A compromised \ac{EV} could therefore grant itself any preferred scope, which makes the user losing control.
We solve this issue by using OAuth's cross-device authorization flow (\rfc{8628}) instead.
Kasselman et al. proposed a security best current practice draft at IETF \cite{I-D.ietf-oauth-cross-device-security} on cross-device flows with OAuth.
However, they only explain common attack vectors with cross-device flows without explicitly suggesting solutions for them.
Instead, they recommend implementers to do formal proof verification of developed concepts before implementing them.
We address this by providing a specific solution by the example of \ac{EV} charging authorization with a formal proof analysis.
The concept may be applied to other applications where the authorized device has displaying capability and a Bluetooth, NFC, or UWB interface, e.g., when authorizing a streaming service on a smart TV.

\section{System-/Adversary Model and Requirements}
\label{sec:model}

We assume an e-mobility system model with \acp{EV}, \acp{CP}, \acp{CPO}, and \acp{EMSP} as described in \cref{sec:emob}. 
Based on this system model and considering relevant threat analyses from related work (cf. \cref{sec:rel:ev}), we derive an adversary model to cover realistic threats to the system in the following sub-section.
Afterwards, we define appropriate security and feasibility requirements, which should enable a solution to address the threats considered while achieving a satisfactory level of usability.

\subsection{Adversary Model}\label{sec:adv}

In this paper, we consider the following kinds of adversaries:
\begin{description}
	\item[Local Adversary.] We consider a local adversary with physical access to devices, specifically, to \acp{EV} and \acp{CP} which are commonly left unattended in public areas for long periods of time. The adversary can physically tamper with the respective devices and extract or modify any local data unless it is tamper protected.
	\item[Network Adversary.] We consider a network hacker according to the commonly used Dolev-Yao model \cite{dolev1983security}. That is, an adversary with full control over network traffic who can arbitrarily send, drop, or modify any network data. However, the adversary cannot break cryptographic primitives unless they have access to the respective private keys.
	\item[Adversary with Access to Leaked Backend Data.] We consider an adversary with access to leaked backend data, i.e., the users' payment credentials. Since in the current e-mobility system, the billing-relevant contract credentials of all users are generated by the respective \acp{EMSP}, this is a serious problem. This opens up the unnecessary threat of a large-scale data leak at an \ac{EMSP}, which would give the adversary access to the billing-relevant data of a large number of users.

\end{description}
\subsection{Requirements}\label{sec:req}

We define the requirements for security ($RS$) and feasibility ($RF$) as follows:

\begin{enumerate}[label=\subscript{RS}{{\arabic*}}, leftmargin=2.5em,itemindent=*]%
	\item \label{sr1}\emph{Secure credential installation.} Mutual authentication of data sent between \ac{CPS} and \ac{EV} for credential installation must be guaranteed. 
	\item \label{sr2}\emph{Fine-grained authorization.} The solution must support authorization constraints, e.g., to define expiration times or charge amount limits for an \ac{EV}'s authorization.
    \item \label{sr3}\emph{Secure key storage.} Private keys must be generated locally in a secure environment (e.g., an \ac{HSM} in the \ac{EV}) and must never leave it. 
	\item \label{sr4}\emph{Crypto-agility.} The system must provide a mechanism to update or replace outdated or broken cryptographic algorithms securely.
	\item \label{sr5}\emph{Secure charge authorization.} The authenticity of charge authorization requests sent from \ac{EV} to \ac{CP} must be guaranteed.
	\item \label{sr6}\emph{Charge data authenticity.} The authenticity of charge data received as attested by the \ac{EV} must be guaranteed towards the \ac{EMSP}.
\end{enumerate}

\begin{enumerate}[label=\subscript{RF}{{\arabic*}}, leftmargin=2.5em,itemindent=*]%
    \item \label{fr1}\emph{Minor overhead.} The system should keep additional communication and computational overhead within the constraints of existing standards.
    \item \label{fr2}\emph{Simple integration.}
    The system should not alter the message flow of existing protocols and only introduce minor changes to message content.
    \item \label{fr3}\emph{Improved usability.} 
    The usability of the proposed system should be improved by utilizing widely used standardized solutions.
\end{enumerate}

\section{Concept}
\label{sec:concept}
This section describes the general concept of the proposed protocol which adds new components to the \ac{PnC} architecture and adds new demands to existing components.
\begin{enumerate*}[label=\emph{(\roman*)}]
    \item The \ac{OEM} must provide a \ac{UA} which is a mobile app running on the driver's smartphone to represent the \ac{RO} in the OAuth cross-device authorization flow.
    It may be integrated into an existing companion app such as most \acp{OEM} already provide for their \acp{EV}.
    \item The \ac{EV} must have an active Internet connection during the installation. Alternatively, the \ac{EV}'s communication can be tunneled through the \ac{CP} using the \ac{VAS} feature of ISO~15118, which is designed to enable extensibility of the standard and already enables the forwarding of traffic between \ac{EV} and Internet.
    \item The \ac{EV} must also be capable of displaying at least eight characters and allow Bluetooth connections to the \ac{UA} which is often implemented to use a smartphone as audio source.
    \item The \ac{EMSP} must provide an OAuth \ac{AS} for customer login and its Contract Certificate signing server must implement an OAuth \ac{RS}.
\end{enumerate*}
OAuth was chosen for its huge popularity in modern identity and access solutions which causes compatibility with existing backend services and a well-known user experience.
Furthermore, OAuth comes with a wide range of field-tested extensions which solve specific issues with existing standards.

\begin{figure}
    \centering
    \includegraphics[width=1.0\linewidth]{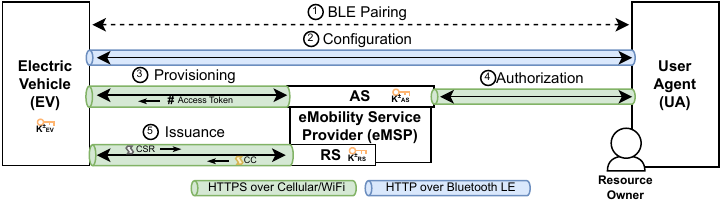}
    \caption{High-level overview of protocol flow.}
    \label{fig:overview}
\end{figure}

\fig{overview} shows a high-level overview of these components; their high-level interactions are as follows.
\begin{enumerate}
    \item The driver (i.e., \ac{RO}), pairs the trusted \ac{UA} with the \ac{EV} via \ac{BLE}.
    \item The \ac{EV} sends its supported \acp{EMSP} via HTTP over an encrypted and authenticated \ac{BLE} connection to the \ac{UA}.
    The \ac{RO} selects its preferred \ac{EMSP}, adjusts authorization details, and sends this configuration via HTTP over \ac{BLE} to the \ac{EV}.
    \item The \ac{EV} forwards these authorization details via HTTPS through the Internet to the selected \ac{EMSP}'s \ac{AS} by initiating the OAuth cross-device authorization flow.
    \item The \ac{RO} then signs in to the \ac{EMSP}'s \ac{AS} on the \ac{UA} via HTTPS over a cellular or WiFi Internet connection and checks the authorization details.
    When granted, the \ac{AS} issues an \acf{AT} to the \ac{EV}, which is signed with its private key $K^-_{AS}$.
    \item This \ac{AT} authorizes the \ac{EV} to request contract certificates from the \ac{EMSP}'s certificate authority which implements an OAuth \acf{RS}.
    Therefore, the \ac{EV} generates a key pair \{$K^+_{EV}$, $K^-_{EV}$\} in its \ac{HSM}.
    Afterwards, the \ac{EV} sends a certificate signing request (CSR) for its public key $K^+_{EV}$, signed with the corresponding private key $K^-_{EV}$, to the \ac{RS}.
    The \ac{RS} verifies the signature using the contained public key $K^+_{EV}$ and responds with a contract certificate (CC) containing the \ac{EV}'s public key $K^+_{EV}$, signed with the \ac{EMSP}'s private key $K^-_{RS}$.
    The \ac{EV} can then use the CC and $K^-_{EV}$ for \ac{PnC} authentication to a \acf{CP}.
\end{enumerate}

\subsection{Connection Establishment}
\begin{figure}[h]
    \centering
    \includegraphics[width=1.0\columnwidth]{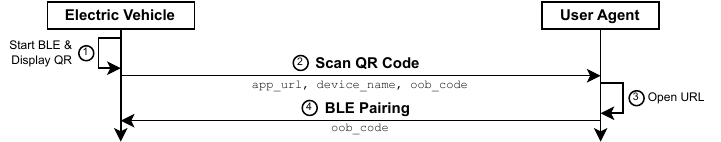}
    \caption{Step 1: \ac{UA} connects to \ac{EV} via \ac{BLE}.}
    \label{fig:connect_ble}
\end{figure}

Before the protocol flow starts, the \ac{UA} connects to the \ac{EV} via \ac{BLE} using out-of-band (OOB) authentication as depicted in \fig{connect_ble}.
To establish the connection easily, the \ac{EV} displays a QR code referring to the user agent, e.g., with a URL like \texttt{https://ua.example.com/connnect?\allowbreak{}device\_name=MyEV\#oob\_code=1234}.
The URL includes the \ac{EV}'s Bluetooth device which the user agent connects to.
Since the \ac{EV} displays the QR code directly on its infotainment screen, typical quishing attacks on QR codes do not apply here unless the \ac{EV} itself is compromised.
To mitigate \acf{MitM} attacks in the Bluetooth connection, out-of-band (OOB) authentication is used, which has been proven by Cäsar et al. \cite{Caesar2022} to be safe.
After pairing, the \ac{UA} connects to the \ac{BLE} HTTP Proxy Service \cite{bluetooth_http_proxy_service} on the \ac{EV}.
This allows the \ac{UA} to communicate with the \ac{EV} via HTTP over the encrypted Bluetooth connection.

\subsection{Configuration Request}
\begin{figure}[h]
    \centering
    \includegraphics[width=1.0\columnwidth]{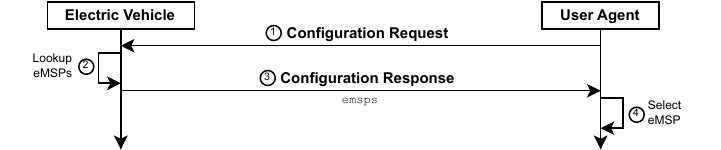}
    \caption{Step 2: \ac{UA} requests configuration from \ac{EV}.}
    \label{fig:config_request}
\end{figure}

After establishing the \ac{BLE} connection, the \ac{UA} looks up a list of \acp{EMSP} pre-configured by the \ac{EV} manufacturer, of which the user selects one.
This process is illustrated in \fig{config_request}.

\subsection{Contract Provisioning Request}\label{sec:concept:CProvReq}
\begin{figure}[h]
    \centering
    \includegraphics[width=1.0\columnwidth]{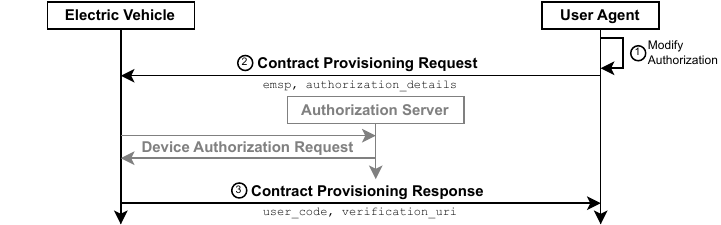}
    \caption{Step 3: \ac{UA} sends a contract provisioning request to \ac{EV}.}
    \label{fig:contract_provisioning_request}
\end{figure}

In contrast to classic \acf{PnC} provisioning, OAuth 2 allows more fine-grained authorizations using the \acf{RAR} standard defined in \rfc{9396}.
This allows the \ac{RO} to specify authorization details on the \ac{UA}, which may contain but are not limited to the following configuration parameters:
\begin{description}
    \item[Authorization Period.] This defines the start and end time within which the \ac{EV} is authorized to charge, e.g., to limit the authorization to the rental period of a rental car.
    \item[Maximum period expanses.] This limits the charging costs within the authorization period to prevent car thieves from charging infinitely.
    \item[Maximum expanses per day.] This limits the charging costs per day to prevent a hijacked contract certificate of a compact car from being effectively used for a truck.
\end{description}
This protocol flow is illustrated in \fig{contract_provisioning_request}.

First, the \ac{RO} modifies the authorization details on the \ac{UA} (1) which sends them together with the selected \ac{EMSP} to the \ac{EV} (2).
Then, the \ac{EV} sends a Device Authorization Request to the \ac{EV} (grayed out, see \sect{device_authorization}).
Finally, the \ac{EV} responds to the \ac{UA} with the user code and the verification URI obtained from the \ac{UA} in the Device Authorization Request.

\subsection{Device Authorization Request}
\label{sec:device_authorization}
\begin{figure}[h]
    \centering
    \includegraphics[width=1.0\linewidth]{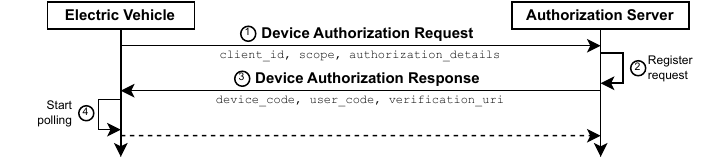}
    \caption{Step 4: \ac{EV} sends a device authorization request to \ac{EMSP}'s \ac{AS}.}
    \label{fig:device_authorization_request}
\end{figure}

The \ac{EV} initiates the Device Authorization Flow with a Device Authorization Request as standardized in \rfc{8628}.
To enable fine-grained authorization, this authorization request contains the authorization details obtained in the previous step from the \ac{UA} to implement a \ac{RAR} as standardized in \rfc{9396}.
The request is illustrated in \fig{device_authorization_request}.

First, the \ac{EV} sends the Device Authorization Request directly via a secure HTTPS request through the Internet to the \ac{EMSP}'s \ac{AS} (1).
The request additionally contains the \ac{EV}'s client ID which was pre-registered by the \ac{OEM}, and the scope required to request contract certificates.
The \ac{EMSP}'s \ac{AS} then registers this request (2), responds with a verification URI, and issues a device code and a user code to the \ac{EV} (3).
The \ac{EV} then starts polling Device Access Token Requests from the \ac{EMSP}'s \ac{AS} (4) and displays the transmitted user code on its infotainment display.

\subsection{Authorization Request}
\label{sec:authorization_request}
\begin{figure}[h]
    \centering
    \includegraphics[width=1.0\columnwidth]{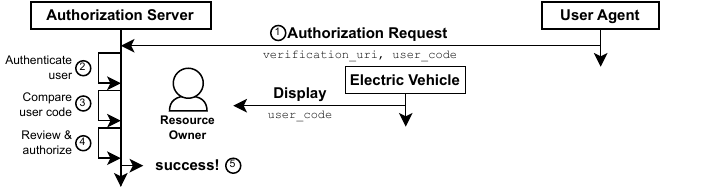}
    \caption{Step 5: \ac{UA} sends Authorization Request to \ac{EMSP}'s \ac{AS}.}
    \label{fig:authorization_request}
\end{figure}

The \ac{RO} must now authorize the \ac{EV} on the \ac{UA} as illustrated in \fig{authorization_request}.

\subsubsection{Authorization Flow}
First, the \ac{UA} opens the verification URI with the user code as query parameter at the \ac{EMSP}'s \ac{AS} in the driver's Web browser (1).
The \ac{EMSP}'s \ac{AS} then authenticates the driver (2) either with an \ac{EMSP} account, or with an \ac{OIDC}-based third-party login (see \ac{OIDC} Extension below).
After successful authentication, the \ac{AS} displays the user code from the query parameter which was originally transferred to the \ac{UA} via Bluetooth.
The driver then compares this user code to the user code that the \ac{EV} displays on its infotainment system (3).
Verifying that both user codes are identical introduces an additional layer of \ac{MitM} protection, as well as a proximity check for the cross-device flow, utilizing the \ac{EV}'s display as an out-of-band channel to the spatially constrained Bluetooth channel.
If the smartphone and the \ac{EV} both had an NFC or UWB interface, these technologies could be used instead of Bluetooth.
When the driver confirms equality of the user codes, the \ac{AS} presents the scope and the authorization details from the \ac{EV}'s device authorization request.
The driver reviews the scope and authorization details (4) to ensure that the \ac{EV} did not manipulate them.
If the \ac{RO} confirms, the \ac{AS} responds with a success message (5).

\subsubsection{OIDC Extension}
\label{sec:concept_oidc}
In step 2, the \ac{EMSP} could also allow the \ac{RO} to login with an existing \acf{OIDC} account at a third party \ac{IdP}, e.g., Apple or Google.
This enables a driver to become a new customer of the chosen \ac{EMSP} without creating a new account.
Instead, a drivers logs in at the \ac{AS} with an \ac{IdP} account, commonly known as the \enquote{Sign in with Apple/Google} button.
If the \ac{IdP} is a payment provider, e.g., Apple Pay, Google Pay, or PayPal, the driver can authorize payment transactions in the login process, comparable to the \enquote{Express Checkout} button in online shops.
Afterwards, the driver continues with step 3.
Since this extension is optional and the underlying technology is straightforward, we placed the technical explanation for interested readers in Appendix \ref{sec:oidc_extension}.

\subsection{Device Access Token Request}
\begin{figure}
    \centering
    \includegraphics[width=1.0\linewidth]{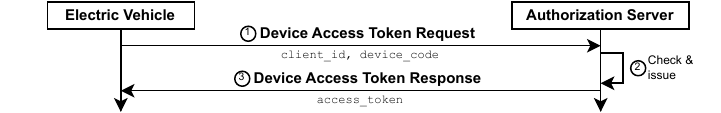}
    \caption{Step 6: \ac{EV} successfully polls a Device Access Token Request from the \ac{AS}.}
    \label{fig:token_request}
\end{figure}

After receiving the Device Authorization Response, the \ac{EV} continuously polls the device access token request illustrated in \fig{token_request}.
The \ac{EV} sends the request through the Internet to the \ac{AS} via HTTPS (1), containing the \ac{EV}'s client ID and the device code from the device authorization response.
The \ac{AS} checks whether the \ac{RO} has already granted the request with the corresponding user code (2).
If granted, the \ac{AS} issues an access token (2) containing the granted scope and authorization details in the Device Access Token Response (3).

\subsection{Contract Certificate Request}
\label{sec:ccr}
\begin{figure}
    \centering
    \includegraphics[width=1.0\linewidth]{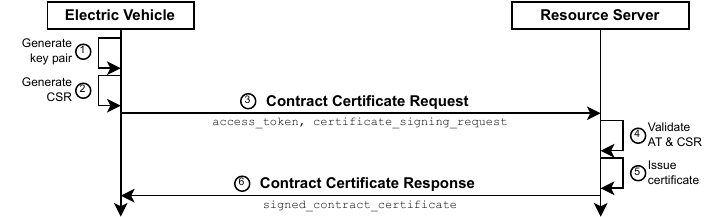}
    \caption{Step 7: \ac{EV} uses access token for a contract certificate request from  \ac{RS}.}
    \label{fig:certificate_request}
\end{figure}

With the access token from the \ac{AS}, the \ac{EV} is authorized to request a contract certificate from the \ac{EMSP}'s \ac{RS} as illustrated in \fig{certificate_request}.
First, the \ac{EV} generates an asymmetric key pair (1) in its \ac{HSM} and generates a \acf{CSR} for it (2). 
The \ac{EV} sends a contract certificate request to the \ac{EMSP}'s \ac{RS} (3), containing the \ac{CSR} and the access token.
The \ac{RS} validates the \ac{CSR} and the access token (4) which must contain a sufficient scope.
If valid, the \ac{RS} signs a new X.509 contract certificate (5) with the \ac{EMSP}'s private key.
This contract certificate contains the \ac{EV}'s public key from the \ac{CSR} and the validity period from the authorization details.
Other authorization details may be represented in the certificate as custom attributes.
The \ac{RS} issues this contract certificate (6) to the \ac{EV} in the response.

\subsection{Using a Contract Certificate}\label{sec:concept:ucc}
After receiving the contract certificate, the \ac{EV} can use its contract credentials (i.e., key pair and corresponding certificate) for \ac{PnC} as defined in ISO~15118.
In other words, this OAuth-based \ac{PnC} extension proposal only affects the installation of the \ac{EV}'s contract certificate.
It does not add any additional requirements to the rest of ISO~15118's \ac{PnC} infrastructure. 
The only difference for the \ac{EV} is that signature creation now requires an interaction with its \ac{HSM}. 
More specifically, the \ac{EV} can use its \ac{HSM}-proteced credentials for \ac{PnC} authentication and to sign meter readings at the \ac{CP}, as described in \cref{sec:emob}, without requiring the \ac{CP} to support OAuth in any way. 

\section{Implementation}\label{sec:impl}
\begin{figure}[h]
    \centering
    \includegraphics[width=.7\linewidth]{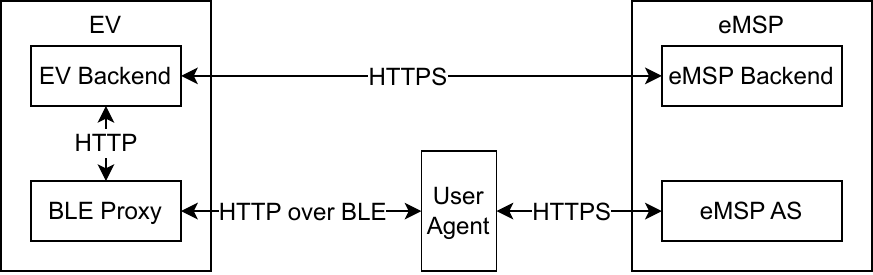}
    \caption{Architecture of the prototype implementation.}
    \label{fig:prototype}
\end{figure}

We implemented the concept described in \sect{concept} and published a prototype on GitHub.\footnote{\label{f:git}\url{https://anonymous.4open.science/r/ev-pnc-oidc-1B84/}}
The architecture of the prototype is depicted in \fig{prototype}. Its components are described in this section.

The \ac{EV} is simulated with a Raspberry Pi 3B with a quad core 1.2 GHz Arm Cortex-A53 CPU which runs a Docker composition \cite{docker_compose} of the required services.
While there is variance among manufacturers, the Pi board can nonetheless be argued to represent a realistic hardware setup \cite{kern2023quantumcharge}, for example, w.r.t. vehicular \acp{ECU} such as an infotainment unit in the Arm Cortex-A performance class~\cite{vehicle-arm} (e.g.,~\cite{ti-infotainment-arm} with a dual Arm Cortex–A15 at 1.2 GHz). 
The \ac{EV} backend implementation is a containerized application developed in Go \cite{go_lang} and a Traefik \cite{traefik} reverse proxy in front.
The \ac{BLE} proxy is a Python \cite{python} script which implements a Bluetooth server with the HTTP Proxy Service \cite{bluetooth_http_proxy_service}.
It proxies HTTP requests from the Bluetooth HTTP Proxy Service via the reverse proxy to the \ac{EV} backend.

The \acf{UA} is an Angular Web application \cite{angular} written in TypeScript \cite{typescript} running on the driver's smartphone.
It utilizes Chromium's Web Bluetooth API \cite{web_bluetooth} to connect the driver's smartphone to the \ac{EV}.

The \acf{EMSP} is a server which runs a Docker composition \cite{docker_compose} of the required services.
The \ac{EMSP} \acf{AS} is a containerized service written in PHP \cite{php} provided by Nginx \cite{nginx} which utilizes the Cloud-based Authlete Authorization Server API \cite{authlete_as}.
The \ac{EMSP} backend is a containerized service which implements the \acf{RS} written in Go \cite{go_lang} that signs the \acf{CSR} using the OpenSSL \cite{openssl} command line interface.
A Traefik \cite{traefik} reverse proxy in front of the \ac{EMSP} backend and \ac{AS} manages TLS certificates and validates the \ac{EV}'s access token when requesting endpoints on the \ac{EMSP} backend.

\section{Evaluation}
\label{sec:eval}
In this section we summarize the results of the performance evaluation, describe the tool-based automated formal security verification, and discuss how the proposed solution can meet the requirements defined in \cref{sec:req}.

\subsection{Performance Evaluation}\label{sec:eval:p}
The goals of the performance evaluation are as follows.
\begin{enumerate*}[label=\emph{(\roman*)}]
	\item Measuring the memory consumption and CPU utilization on the \ac{EV} software to estimate additional hardware requirements induced by this approach.
    \item Measuring the data volume requirements for Internet connections by the \ac{EV} to estimate the additional costs for \acp{OEM}.
	\item Identifying the processing durations of steps without user interactions to estimate the impact of data transfers/processing on the user experience and to identify time-consuming steps.
\end{enumerate*}

The proof-of-concept implementation was evaluated on the following hardware by manually clicking through the setup 50 times.
The \ac{EV} software was thereby running on a Raspberry Pi B3 1 GB with Raspberry Pi OS (Debian 12.7 64-bit), the \acf{UA} was running in Chrome 128 on a Google Pixel 6 with Android 14, and the \ac{EMSP} software was running on a Lenovo ThinkPad T14s G1 Ryzen 5 PRO 4650U with Windows 11 Pro 23H2 in WSL2.

\begin{figure}[h]
    \centering
    \includegraphics[width=1.0\linewidth]{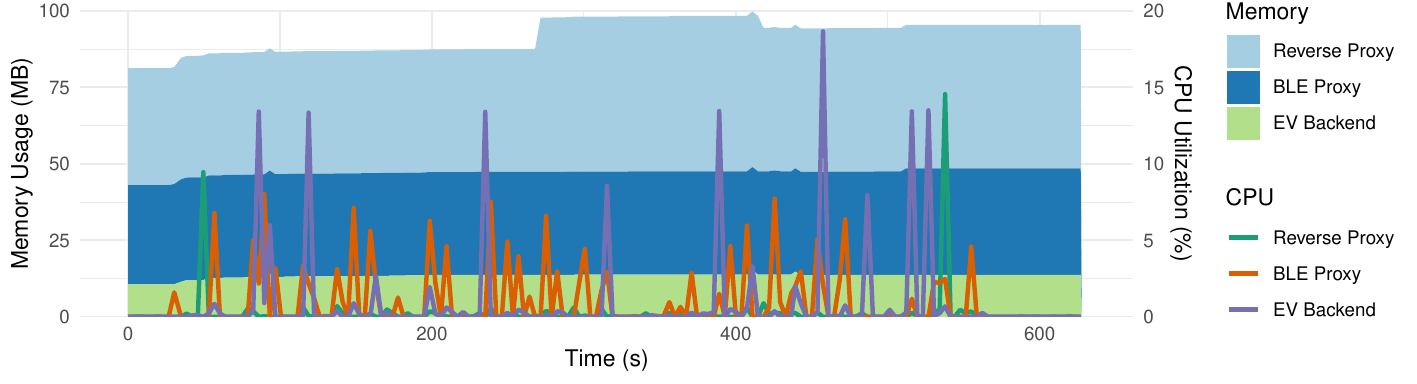}
    \caption{Memory consumption (stacked) and CPU utilization (not stacked) of the \ac{EV} software by services over evaluation time.}
    \label{fig:hardware_stats}
\end{figure}
\begin{table}
    \centering
    \caption{Statistics about memory consumption and CPU utilization of the \ac{EV} software by services over evaluation time.}
    \begin{tabular}{l|lll|lll}
    \multicolumn{1}{c|}{\multirow{2}{*}{\textbf{Service}}} & \multicolumn{3}{c|}{\textbf{Memory (MB)}}  & \multicolumn{3}{c}{\textbf{CPU (\%)}}       \\
    \multicolumn{1}{c|}{}                                  & \textbf{Min} & \textbf{Max} & \textbf{Avg} & \textbf{Min} & \textbf{Max} & \textbf{q.95} \\ \hline
    \textbf{EV backend}                                    & 10.654       & 15.204       & 13.445       & 0.00\%       & 8.03\%       & 5.79\%        \\
    \textbf{BLE proxy}                                     & 32.422       & 34.802       & 33.827       & 0.00\%       & 14.56\%      & 0.53\%        \\
    \textbf{Reverse proxy}                                 & 38.210       & 50.856       & 44.751       & 0.00\%       & 20.76\%      & 10.49\%      
    \end{tabular}
    \label{tab:ev_stats}
\end{table}
The memory consumption and CPU utilization were measured with the \texttt{docker stats} command periodically during the execution.
The results are compiled in \fig{hardware_stats} and Table~\ref{tab:ev_stats}.
We derive a memory requirement of at most 15 MB for the EV backend plus 75 MB for the reverse proxy and BLE proxy combined, if not already implemented.
The four cores of the Raspberry Pi 3B's CPU (max: 400 \%), never exceeded 21 \%, not even by the elliptic curve key generation and signing process with the P521 curve which instead only led to a shorter idle time  of the \ac{EV} backend compared to the BLE proxy, as indicated by their 95\% quantiles compared to their maximum CPU utilization.
Instead, the reverse proxy induced the highest CPU utilization, which indicates a very good performance of the \ac{EV} backend.

To measure the traffic between the \ac{EV} backend and the \ac{EMSP} \ac{AS} and backend which must be transferred through the Internet, we used tcpdump to dump the Pi's traffic over network interface eth0, and filtered it with Wireshark.
The total traffic was 434,297 bytes which corresponds to an average data volume of 8,686 bytes per authorization.
This includes the HTTPS communication including TCP, IP, and Ethernet headers as well as TLS handshakes, but excludes corresponding communications to other parties such as DNS or ARP requests, certificate validity verifications or telemetry data transfers.
However, because of TLS~1.3's 0-RTT feature, the TLS handshakes to the \ac{AS} and the \ac{RS} were omitted after the first communication.
The polling mechanism of the Device Authorization Flow might increase the traffic slightly.
This additional data volume required by the OAuth extensions is negligible compared to the fact that the HTML body of \texttt{www.google.com} is about 25 kilobytes long.

\begin{figure}[h]
    \centering
    \includegraphics[width=1.0\linewidth]{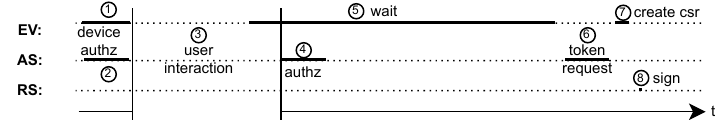}
    \caption{Evaluation of processing times.}
    \label{fig:evaluation}
\end{figure}
We measured the execution times of relevant processing steps on the \ac{EV}, the \ac{AS}, and the \ac{RS} which do not involve user interaction.
\fig{evaluation} illustrates the results.
The measurements start with the device authorization (1; avg. 0.468 s) on the \ac{EV} after it received the authorization details.
Almost all of the execution time is caused by the \ac{AS} (2; avg. avg. 0.449 s).
After the user authenticated and authorized the \ac{EV} (3), the \ac{AS} processes the authorization response (4; avg. 0.449 s).
After waiting for 5 seconds (5), the \ac{EV} sends the token request to the \ac{AS} (6; 0.431 s) to obtain a valid token before it starts generating the asymmetric key pair and creating the certificate signing request (7; avg. 0.140 s).
Signing the contract certificate (8; avg. 0.026 s) is faster because the ThinkPad has more CPU power, and it does not need to generate a key pair.

This shows that our implementation performs well even on weak hardware of the Raspberry Pi 3B and the performance overhead induced by the issuance of the contract certificate by the \ac{EMSP} backend is negligible.
The most significant time impacts are waiting times for the user to interact with the \ac{AS}.
The relatively large processing times of the \ac{AS} are caused by the network latency to the Authlete Cloud API which is caused by the average ping of 130 milliseconds in the testing environment.
In conclusion, it can be stated that the proposed solution performs well and causes minimal overhead.

\subsection{Formal Security Verification}\label{sec:formal}
We analyze the security of the proposed protocol in the symbolic model, also called the Dolev-Yao model \cite{dolev1983security}. Hereby, the adversary has full control over the network. Additionally, cryptographic primitives are assumed to be perfectly secure, i.e., cannot be broken without knowing the respective private key.
Specifically, the formal security analysis focuses on the authentication properties of the OAuth-based credential installation protocol (\ref{sr1}). Other security requirements are discussed in \cref{sec:eval:dis}.

The symbolic security verification is performed with the Tamarin prover \cite{meier2013tamarin}.
Tamarin is a state-of-the-art tool for automated security protocol analysis.
With Tamarin, a protocol is modeled via a set of rules, defining the communication and data processing steps.
Additionally, security properties are modeled via lemmas that must hold for all possible execution traces of the model.
Afterwards, Tamarin starts in a state where the lemma was violated and performs an exhaustive backwards search.
If a valid execution trace is found that violates the lemma, this trace serves as a counterexample, i.e., a possible attack. Otherwise, the security property is shown to hold.

For the verification of strong authentication properties, we adopt the commonly used notion of injective agreement:
\begin{definition}[Injective agreement] \label{def:inj-agreement}
A protocol guarantees \emph{injective agreement} to an honest initiator $A$ with an honest responder $B$ on a set of data items $\vec{ds}$ if, whenever $A$, acting as initiator, completes a run with the protocol,
apparently with responder $B$, then $B$ has previously been running the protocol, apparently with $A$, and $B$ was acting as a responder in this run.
Moreover, each run of $A$ corresponds to a unique run of $B$ and both agents agree on the values of the variables in $\vec{ds}$.~\cite{lowe1997hierarchy}%
\end{definition}

For our Tamarin model, we assume that initial exchange of \ac{OOB} data (via QR code) for Bluetooth pairing is secure (i.e., modeled via a secure channel). Additionally, 
any of the user's interactions with their user agent are assumed to be secure.
Furthermore, any long-term secrets (e.g., private keys of passwords) of actors involved in a specific protocol run are assumed to be secure.
Hence, to keep the needed assumptions as weak as possible, any actors that are not directly involved in a specific protocol run can be compromised.
For example, the security of a protocol run with \ac{EMSP} \emph{A} should not be compromised by a key leak at \ac{EMSP} \emph{B}.
The establishment of a shared symmetric key during Bluetooth pairing is modeled as per the most recent Bluetooth core specification \cite{bt-core54}.
The establishment of symmetric keys during the \ac{TLS} handshakes are modeled as key encapsulations based on the \ac{EMSP} public key for simplicity.
This includes the \ac{TLS} handshakes between user agent and \ac{EMSP} as well as \ac{EV} and \ac{EMSP}; both with unilateral authentication by the \ac{EMSP}.
Any other communication and data processing steps of the proposed protocol are modeled as described in \cref{sec:concept}.
Our full Tamarin model definition can be found online (cf. \cref{f:git} on \cpageref{f:git}).

As mentioned, we mainly focus on the verification of strong authentication properties for the installation of contract credentials based on the notion of injective agreement.
This notion can be transferred to our use case in the form of a Tamarin lemma as shown by the example \cref{fig:eval-inj} in \cref{sec:app:tam}.

Besides the injective agreement lemmas to prove secure authentication, the model also includes verifications of intermediate steps and weaker security properties based on Lowe's hierarchy of authentication properties \cite{lowe1997hierarchy}.
The model further includes several correctness lemmas, i.e., lemmas that verify that the correct/expected execution of the model is possible to avoid false security proofs by an erroneous model.
Additionally, we include several performance optimizations for the automated proofs to terminate within an acceptable timeframe, such as reuse lemmas (which can provide partial proofs for reuse in later lemmas) and an oracle (which guides the automated proof execution, affecting runtime but not correctness).

The automated symbolic analysis with Tamarin and the model shows that the proposed protocol satisfies the security properties, as no counterexamples are found. 
Specifically, the analysis shows that injective agreement over the contract credentials holds between \ac{EV} and \ac{EMSP} during credential installation. Injective agreement is shown to hold in both directions, i.e., installation request and response.
Results of the automated symbolic analysis and details on how to recreate the evaluations results with the Tamarin prover can be found online (cf. \cref{f:git} on \cpageref{f:git}).
In short, on a standard laptop, the verification of all lemmas with automatic sources generation finished in roughly 228 minutes and consisted of a total of $131,305$ proof steps.

\subsection{Requirement Discussion}
\label{sec:eval:dis}
In the following, we discuss how the proposed solution can address the requirements for security ($RS$) and feasibility ($RF$) from \sect{req}. 
An overview is provided in \cref{tab:con:prot:perf}. 
Starting with the security requirements, secure credential installation \ref{sr1} can be ensured as shown in the formal verification from \cref{sec:formal}.
Fine-grained authorization \ref{sr2} is supported by the usage of \acp{RAR} as discussed in \cref{sec:concept:CProvReq}.
Secure key storage \ref{sr3} is enabled by the usage of an \ac{HSM} in the \ac{EV} (cf. \cref{sec:ccr}).
Since the proposed solution does not rely on specific algorithms, crypto-agility \ref{sr4} is given because the OAuth framework already enables it by interchangeable state of the art crypto algorithms.
Additionally, the proposed solution supports secure charge authorization \ref{sr5} and charge data authenticity \ref{sr6} with the default ISO~15118 processes.
For formal/provable security in \ref{sr5} and \ref{sr6}, existing approaches can be used as discussed in \cref{sec:concept:ucc}. 

\begin{table*}[h]
	\centering
	\caption{Differences to the Existing Solution}
	\label{tab:con:prot:perf}
	\resizebox{\linewidth}{!}{%
		\begin{tabular}{l@{\hspace{1em}} p{16em}@{\hspace{1em}} p{16em}}
			\toprule
			\textbf{Requirement} & \textbf{Our Method}      & \textbf{ISO 15118} \\ \midrule
			\ref{sr1} & Formally Verified Injective Agreement Properties & Unverified Security Claims \\ 
			\ref{sr2} & Fine-Grained Authorization & Binary Authorization \\ 
			\ref{sr3} & Secure Key Storage via \ac{HSM} & Secure Key Storage Optional \\ 
			\ref{sr4} & Crypto-Agility Possible & Fixed Cryptographic Algorithms \\ 
			\ref{sr5} & Charge Authorization via Secure Contract Credentials & Charge Authorization via Contract Credentials \\ 
			\ref{sr6} & Charge Data Authenticity via Secure Contract Credentials & Charge Data Authenticity via Contract Credentials \\ 
            
			\ref{fr1} & Additional Overhead Evaluated in \cref{sec:eval:p} & N/A \\ 
			\ref{fr2} & Credential Installation Using Default OAuth Methods & Credential Installation Requires Proprietary Solutions \\ 
			\ref{fr3} & App-Guided Setup Flow & Proprietary Setup Flow \\ \bottomrule
		\end{tabular}%
	}
\end{table*}

Regarding feasibility requirements, the proposed solution only generated minor overhead \ref{fr1} as shown in \cref{sec:eval:p}.
We argue that the simple integration requirement \ref{fr2} is met since the OAuth-based credential installation process works with default OAuth methods and after installation, credentials can be used as usual for ISO~15118 (cf. \cref{sec:concept:ucc}).
Additionally, since the installation traffic can be tunneled via an ISO~15118 \ac{VAS}, the requirements for new message flow paths can be kept to a minimum.
Finally, we argue that the proposed solution provides a high level of usability \ref{fr3} because it introduces an app-guided setup flow based on open standards which is similar for each combination of \ac{EV} model and \ac{EMSP}.
This flow uses familiar interaction concepts like scanning a QR code, choosing an \ac{EMSP} from a list, logging in with an existing \ac{EMSP} account, comparing security numbers on two screens, and authorizing an OAuth client.
Instead of the \ac{EMSP} account, drivers can also login with \ac{OIDC} through a payment provider like when using the express checkout feature in an online shop.
Furthermore, the proposed solution can obsolete some of the ISO~15118 \ac{PKI} requirements (cf. \cref{fig:isoPKI}) since it does not need the \ac{OEM} and \ac{CPS} certificate chains.
Essentially, the \ac{OEM} chain is replaced by the user's OAuth authentication and the \ac{CPS} chain by the OAuth-based contract credential installation.


\section{Conclusion}
\label{sec:conclusion}
To enhance the usability of \acp{EV}, the \ac{PnC} feature of ISO 15118 plays a critical role. However, a key criticism of PnC lies in its complex and poorly defined \ac{PKI} and credential installation/management processes.
In this paper, we present a method to address these challenges by proposing a solution for \ac{EV} credential installation based on the widely adopted OAuth 2.0 standard. This approach improves usability by introducing an app-guided installation process, which operates seamlessly across various combinations of \acp{EMSP} and \acp{EV}, leveraging familiar user interaction patterns. Additionally, the solution enables users to easily become customers of different \acp{EMSP} by simplifying the contract credential installation process, which operates seamlessly across various combinations of by utilizing \ac{OIDC}-based payment providers, facilitating an express checkout process similar to that found in online retail environments. 
Security is further enhanced through fine-grained authorizations enabled by OAuth’s \acfp{RAR}. Through a proof-of-concept implementation, we demonstrate that the proposed solution meets essential performance and usability requirements. Moreover, security properties were formally verified using the Tamarin prover, ensuring that the solution satisfies critical security criteria. This verification underscores the robustness of the OAuth cross-device authorization flow, which has been applied to the secure installation of billing-relevant contract credentials for \acp{EV}.
Lastly, the implementation's source code and the Tamarin models used for verification are made available online, ensuring the reproducibility of the presented results and providing a foundation for future applications in similar contexts.

%
\section*{Acknowledgment}

We want to thank Authlete Inc. for providing their Authorization Server API for testing purposes to us for free.

This work was supported in part by Deutsche Forschungsgemeinschaft (DFG, German Research Foundation), project number 503329135, and under Grant ME2727/3-1, and in part by the
 Open Access Publishing Fund of the University of Tübingen, Germany.

%
\bibliographystyle{splncs04}
\bibliography{literature}

\begin{thebibliography}{10}
\providecommand{\url}[1]{#1}
\csname url@samestyle\endcsname
\providecommand{\newblock}{\relax}
\providecommand{\bibinfo}[2]{#2}
\providecommand{\BIBentrySTDinterwordspacing}{\spaceskip=0pt\relax}
\providecommand{\BIBentryALTinterwordstretchfactor}{4}
\providecommand{\BIBentryALTinterwordspacing}{\spaceskip=\fontdimen2\font plus
\BIBentryALTinterwordstretchfactor\fontdimen3\font minus
  \fontdimen4\font\relax}
\providecommand{\BIBforeignlanguage}[2]{{%
\expandafter\ifx\csname l@#1\endcsname\relax
\typeout{** WARNING: IEEEtran.bst: No hyphenation pattern has been}%
\typeout{** loaded for the language `#1'. Using the pattern for}%
\typeout{** the default language instead.}%
\else
\language=\csname l@#1\endcsname
\fi
#2}}
\providecommand{\BIBdecl}{\relax}
\BIBdecl

\bibitem{iso2}
ISO/IEC, ``{Road Vehicles -- Vehicle-to-Grid Communication Interface -- Part 2:
  Network and Application Protocol Requirements},'' ISO, ISO 15118-2:2014,
  2014.

\bibitem{isoPkiConsiderations}
\BIBentryALTinterwordspacing
ChargePoint, DigiCert, and Eonti, ``{Practical Considerations for
  Implementation and Scaling ISO 15118 into a Secure EV Charging Ecosystem},''
  May 2019. [Online]. Available:
  \url{https://www.chargepoint.com/files/15118whitepaper.pdf}
\BIBentrySTDinterwordspacing

\bibitem{ELAADNL2018}
\BIBentryALTinterwordspacing
{ElaadNL}, ``{Exploring the Public Key Infrastructure for ISO 15118 in the EV
  Charging Ecosystem},'' Arnhem, The Netherlands, November 2018. [Online].
  Available:
  \url{https://www.elaad.nl/news/publication-exploring-the-public-key-infrastructure-for-iso-15118-in-the-ev-charging-ecosystem/}
\BIBentrySTDinterwordspacing

\bibitem{vde-ar-2802}
{VDE}, ``{Handling of Certificates for Electric Vehicles, Charging
  Infrastructure and Backend Systems within the Framework of ISO 15118},''
  Verband der Elektrotechnik Elektronik Informationstechnik e.V., VDE-AR-E
  2802-100-1:2019-12, 12 2019.

\bibitem{meier2013tamarin}
S.~Meier, B.~Schmidt, C.~Cremers, and D.~Basin, ``{The TAMARIN Prover for the
  Symbolic Analysis of Security Protocols},'' in \emph{Computer Aided
  Verification}, ser. LNCS, vol. 8044.\hskip 1em plus 0.5em minus 0.4em\relax
  Springer, 2013, pp. 696--701.

\bibitem{RFC6749}
\BIBentryALTinterwordspacing
D.~Hardt, ``{The OAuth 2.0 Authorization Framework},'' Internet Requests for
  Comments, RFC Editor, RFC 6749, October 2012. [Online]. Available:
  \url{https://www.rfc-editor.org/rfc/rfc6749.txt}
\BIBentrySTDinterwordspacing

\bibitem{rfc7231}
\BIBentryALTinterwordspacing
R.~T. Fielding and J.~Reschke, ``{Hypertext Transfer Protocol (HTTP/1.1):
  Semantics and Content},'' Internet Requests for Comments, RFC Editor, RFC
  7231, June 2014. [Online]. Available:
  \url{https://www.rfc-editor.org/rfc/rfc7231.txt}
\BIBentrySTDinterwordspacing

\bibitem{rfc8259}
\BIBentryALTinterwordspacing
T.~Bray, ``{The JavaScript Object Notation (JSON) Data Interchange Format},''
  Internet Requests for Comments, RFC Editor, RFC 8259, October 2017. [Online].
  Available: \url{https://www.rfc-editor.org/rfc/rfc8259.txt}
\BIBentrySTDinterwordspacing

\bibitem{RFC8628}
\BIBentryALTinterwordspacing
W.~Denniss, J.~Bradley, M.~Jones, and H.~Tschofenig, ``{OAuth 2.0 Device
  Authorization Grant},'' Internet Requests for Comments, RFC Editor, RFC 8628,
  August 2019. [Online]. Available:
  \url{https://www.rfc-editor.org/rfc/rfc8628.txt}
\BIBentrySTDinterwordspacing

\bibitem{RFC9396}
T.~Lodderstedt, J.~Richer, and B.~Campbell, ``{OAuth 2.0 Rich Authorization
  Requests},'' Internet Requests for Comments, RFC Editor, RFC 9396, May 2023.

\bibitem{plappert2021attack}
C.~Plappert, D.~Zelle, H.~Gadacz, R.~Rieke, D.~Scheuermann, and C.~Krau{\ss},
  ``{Attack Surface Assessment for Cybersecurity Engineering in the Automotive
  Domain},'' in \emph{2021 29th Euromicro Int. Conf. on Parallel, Distributed
  and Network-based Processing (PDP)}.\hskip 1em plus 0.5em minus 0.4em\relax
  IEEE, 2021, pp. 266--275.

\bibitem{antoun2020detailed}
J.~Antoun, M.~E. Kabir, B.~Moussa, R.~Atallah, and C.~Assi, ``{A detailed
  Security Assessment of the EV Charging Ecosystem},'' \emph{IEEE Network},
  vol.~34, no.~3, pp. 200--207, 2020.

\bibitem{falk2012electric}
R.~Falk and S.~Fries, ``{Electric Vehicle Charging Infrastructure Security
  Considerations and Approaches},'' \emph{Proc. of INTERNET}, pp. 58--64, 2012.

\bibitem{fuchs2020securing}
A.~Fuchs, D.~Kern, C.~Krau{\ss}, and M.~Zhdanova, ``{Securing Electric Vehicle
  Charging Systems through Component Binding},'' in \emph{Computer Safety,
  Reliability, and Security: 39th International Conference, SAFECOMP 2020,
  Lisbon, Portugal, September 16--18, 2020, Proceedings 39}.\hskip 1em plus
  0.5em minus 0.4em\relax Springer, 2020, pp. 387--401.

\bibitem{kern2021analysis}
D.~Kern and C.~Krau{\ss}, ``{Analysis of e-Mobility-based Threats to Power Grid
  Resilience},'' in \emph{Proceedings of the 5th ACM Computer Science in Cars
  Symposium}, 2021, pp. 1--12.

\bibitem{fuchs2020a}
A.~Fuchs, D.~Kern, C.~Krau{\ss}, and M.~Zhdanova, ``{TrustEV: Trustworthy
  Electric Vehicle Charging and Billing},'' in \emph{Proceedings of the 35th
  {ACM/SIGAPP} Symposium on Applied Computing {SAC} 2020}.\hskip 1em plus 0.5em
  minus 0.4em\relax {ACM}, 2020.

\bibitem{fuchs2020hip}
A.~Fuchs, D.~Kern, C.~Krau\ss{}, and M.~Zhdanova, ``{HIP: HSM-Based Identities
  for Plug-and-Charge},'' in \emph{Proceedings of the 15th International
  Conference on Availability, Reliability and Security}, ser. ARES '20.\hskip
  1em plus 0.5em minus 0.4em\relax New York, NY, USA: Association for Computing
  Machinery, 2020.

\bibitem{fuchs2020hip20}
A.~Fuchs, D.~Kern, C.~Krau\ss{}, M.~Zhdanova, and R.~Heddergott, ``{HIP-20:
  Integration of Vehicle-HSM-Generated Credentials into Plug-and-Charge
  Infrastructure},'' in \emph{Computer Science in Cars Symposium}, ser. CSCS
  '20.\hskip 1em plus 0.5em minus 0.4em\relax New York, NY, USA: Association
  for Computing Machinery, 2020.

\bibitem{van2015securing}
F.~van~den Broek, E.~Poll, and B.~Vieira, ``{Securing the Information
  Infrastructure for EV Charging},'' in \emph{{Wireless and Satellite Systems:
  7th Int. Conf., WiSATS 2015, Bradford, UK, July 6-7, 2015. Revised Selected
  Papers 7}}.\hskip 1em plus 0.5em minus 0.4em\relax Springer, 2015, pp.
  61--74.

\bibitem{van2019non}
P.~Van~Aubel, E.~Poll, and J.~Rijneveld, ``{Non-Repudiation and End-to-End
  Security for Electric-Vehicle Charging},'' in \emph{2019 IEEE PES Innovative
  Smart Grid Technologies Europe (ISGT-Europe)}.\hskip 1em plus 0.5em minus
  0.4em\relax IEEE, 2019, pp. 1--5.

\bibitem{kern2023quantumcharge}
D.~Kern, C.~Krau{\ss}, T.~Lauser, N.~Alnahawi, A.~Wiesmaier, and
  R.~Niederhagen, ``{QuantumCharge: Post-Quantum Cryptography for Electric
  Vehicle Charging},'' in \emph{International Conference on Applied
  Cryptography and Network Security}.\hskip 1em plus 0.5em minus 0.4em\relax
  Springer, 2023, pp. 85--111.

\bibitem{Firoozjaei2019EVChain:Charging}
M.~D. Firoozjaei, A.~Ghorbani, H.~Kim, and J.~Song, ``{EVChain: A
  Blockchain-based Credit Sharing in Electric Vehicles Charging},'' in
  \emph{2019 17th International Conference on Privacy, Security and Trust
  (PST)}.\hskip 1em plus 0.5em minus 0.4em\relax IEEE, 8 2019.

\bibitem{Xu2021EVchain:Vehicles}
S.~Xu, X.~Chen, and Y.~He, ``{EVchain: An Anonymous Blockchain-Based System for
  Charging-Connected Electric Vehicles},'' \emph{Tsinghua Science and
  Technology}, vol.~26, no.~6, 12 2021.

\bibitem{knirsch2018privacy}
F.~Knirsch, A.~Unterweger, and D.~Engel, ``{Privacy-Preserving Blockchain-based
  Electric Vehicle Charging with Dynamic Tariff Decisions},'' \emph{Computer
  Science-Research and Development}, vol.~33, no. 1-2, pp. 71--79, 2018.

\bibitem{richter2021exploring}
D.~Richter and J.~Anke, ``{Exploring Potential Impacts of Self-Sovereign
  Identity on Smart Service Systems: An Analysis of Electric Vehicle Charging
  Services},'' in \emph{Business Information Systems}, 2021, pp. 105--116.

\bibitem{Hoess2022WithOW}
A.~Hoess, T.~Roth, J.~Sedlmeir, G.~Fridgen, and A.~Rieger, ``{With or Without
  Blockchain? Towards a Decentralized, SSI-based eRoaming Architecture},'' in
  \emph{Hawaii International Conference on System Sciences}, 2022.

\bibitem{kailus2024self}
A.~Kailus, D.~Kern, and C.~Krau{\ss}, ``{Self-Sovereign Identity for Electric
  Vehicle Charging},'' in \emph{International Conference on Applied
  Cryptography and Network Security}.\hskip 1em plus 0.5em minus 0.4em\relax
  Springer, 2024, pp. 137--162.

\bibitem{li2016portunes}
H.~Li, G.~D{\'a}n, and K.~Nahrstedt, ``{Portunes+: Privacy-Preserving Fast
  Authentication for Dynamic Electric Vehicle Charging},'' \emph{IEEE
  Transactions on Smart Grid}, vol.~8, no.~5, pp. 2305--2313, 2016.

\bibitem{yucel2019efficient}
F.~Yucel, K.~Akkaya, and E.~Bulut, ``{Efficient and Privacy Preserving Supplier
  Matching for Electric Vehicle Charging},'' \emph{Ad Hoc Networks}, vol.~90,
  p. 101730, 2019.

\bibitem{kern2022integrating}
D.~Kern, T.~Lauser, and C.~Krau{\ss}, ``{Integrating Privacy into the Electric
  Vehicle Charging Architecture},'' \emph{Proceedings on Privacy Enhancing
  Technologies}, vol.~3, pp. 140--158, 2022.

\bibitem{zelle2018anonymous}
D.~Zelle, M.~Springer, M.~Zhdanova, and C.~Krau{\ss}, ``{Anonymous Charging and
  Billing of Electric Vehicles},'' in \emph{Proceedings of the 13th
  International Conference on Availability, Reliability and Security}, 2018,
  pp. 1--10.

\bibitem{zhao2015secure}
T.~Zhao, C.~Zhang, L.~Wei, and Y.~Zhang, ``{A Secure and Privacy-Preserving
  Payment System for Electric Vehicles},'' in \emph{2015 IEEE International
  Conference on Communications (ICC)}.\hskip 1em plus 0.5em minus 0.4em\relax
  IEEE, 2015, pp. 7280--7285.

\bibitem{gadacz2021evaluation}
H.~Gadacz, ``{Evaluation of Electric Mobility Authentication Approaches},'' in
  \emph{Proceedings of the 5th ACM Computer Science in Cars Symposium}, 2021,
  pp. 1--10.

\bibitem{timpner2013secure}
J.~Timpner, D.~Sch{\"u}rmann, and L.~Wolf, ``{Secure Smartphone-based
  Registration and Key Deployment for Vehicle-to-Cloud Communications},'' in
  \emph{Proceedings of the ACM Workshop on Security, Privacy \& Dependability
  for Cyber Vehicles}, 2013, pp. 31--36.

\bibitem{ciba}
\BIBentryALTinterwordspacing
G.~Fernandez, F.~Walter, A.~Nennker, D.~Tonge, and B.~Campbell, ``{OpenID
  Connect Client-Initiated Backchannel Authentication Flow - Core 1.0},''
  OpenID Foundation, Tech. Rep., September 2021. [Online]. Available:
  \url{https://openid.net/specs/openid-client-initiated-backchannel-authentication-core-1\_0.html}
\BIBentrySTDinterwordspacing

\bibitem{I-D.ietf-oauth-cross-device-security}
P.~Kasselman, D.~Fett, and F.~Skokan, ``{Cross-Device Flows: Security Best
  Current Practice},'' Working Draft, IETF Secretariat, Internet-Draft
  draft-ietf-oauth-cross-device-security-08, July 2024.

\bibitem{dolev1983security}
D.~Dolev and A.~Yao, ``{On the Security of Public Key Protocols},'' \emph{IEEE
  Transactions on information theory}, vol.~29, no.~2, pp. 198--208, 1983.

\bibitem{Caesar2022}
\BIBentryALTinterwordspacing
M.~Cäsar, T.~Pawelke, J.~Steffan, and G.~Terhorst, ``{A Survey on Bluetooth
  Low Energy Security and Privacy},'' \emph{Computer Networks}, vol. 205, p.
  108712, January 2022. [Online]. Available:
  \url{https://www.sciencedirect.com/science/article/pii/S1389128621005697}
\BIBentrySTDinterwordspacing

\bibitem{bluetooth_http_proxy_service}
\BIBentryALTinterwordspacing
R.~Heydon, K.~Kerai, M.~Olsson, M.~Holtmann, M.~Andersson, N.~Granqvist,
  F.~Berntsen, C.~Hansen, V.~Zhodzishsky, S.~Walsh, J.~Decuir, K.~Shingala,
  J.~Gros, Y.~Kwon, and M.~Masuda, ``{HTTP Proxy Service 1.0},'' {Bluetooth
  SIG}, Tech. Rep., 2015. [Online]. Available:
  \url{https://www.bluetooth.org/docman/handlers/downloaddoc.ashx?doc\_id=308344}
\BIBentrySTDinterwordspacing

\bibitem{docker_compose}
{Docker Inc.}, ``{Docker Compose Overview},''
  \url{https://docs.docker.com/compose/}, accessed: 2024-08-08.

\bibitem{vehicle-arm}
\BIBentryALTinterwordspacing
{ARM Inc.}, ``{A Starter's Guide to ARM Processing Power in Automotive},''
  2018. [Online]. Available:
  \url{https://community.arm.com/arm-community-blogs/b/embedded-blog/posts/a-starters-guide-to-arm-processing-power-in-automotive}
\BIBentrySTDinterwordspacing

\bibitem{ti-infotainment-arm}
\BIBentryALTinterwordspacing
{Texas Instruments}, ``{DRA745} -- {Infotainment Applications Processor},''
  2019. [Online]. Available: \url{https://www.ti.com/product/DRA745}
\BIBentrySTDinterwordspacing

\bibitem{go_lang}
{Google Inc.}, ``{The Go Programming Language},'' \url{https://go.dev/},
  accessed: 2024-08-08.

\bibitem{traefik}
{Traefik Labs}, ``{Traefik Proxy Documentation},''
  \url{https://doc.traefik.io/traefik/}, accessed: 2024-09-07.

\bibitem{python}
{Python Software Foundation}, ``{Welcome to Python.org},''
  \url{https://www.python.org/}, accessed: 2024-08-08.

\bibitem{angular}
{Google Inc.}, ``{Angular Home},'' \url{https://angular.dev/}, accessed:
  2024-08-08.

\bibitem{typescript}
{Microsoft Inc.}, ``{TypeScript: JavaScript with Syntax for Types},''
  \url{https://www.typescriptlang.org/}, accessed: 2024-08-08.

\bibitem{web_bluetooth}
\BIBentryALTinterwordspacing
R.~Grant and O.~Ruiz-Henríquez, ``{Web Bluetooth},''
  \url{https://webbluetoothcg.github.io/web-bluetooth/}, {World Wide Web
  Consortium}, {Living Internet Standard}, July 2024. [Online]. Available:
  \url{https://webbluetoothcg.github.io/web-bluetooth/}
\BIBentrySTDinterwordspacing

\bibitem{php}
{PHP Documentation Group}, ``{What is PHP?}''
  \url{https://www.php.net/manual/en/intro-whatis.php}, accessed: 2024-08-08.

\bibitem{nginx}
{F5 Inc.}, ``{Nginx},'' \url{https://nginx.org/en/}, accessed: 2024-08-08.

\bibitem{authlete_as}
{Authlete Inc.}, ``{Authlete Overview},''
  \url{https://www.authlete.com/developers/overview/}, accessed: 2024-08-08.

\bibitem{openssl}
{OpenSSL Foundation}, ``{OpenSSL},'' \url{https://www.openssl.org/}, accessed:
  2024-08-08.

\bibitem{lowe1997hierarchy}
G.~Lowe, ``{A Hierarchy of Authentication Specifications},'' in \emph{Computer
  Security Foundations Workshop}.\hskip 1em plus 0.5em minus 0.4em\relax IEEE,
  1997, pp. 31--43.

\bibitem{bt-core54}
{Bluetooth SIG}, ``{Bluetooth Core Specification},'' Core Specification Working
  Group, Tech. Rep. v5.4, 2023.

\bibitem{OidcCore}
\BIBentryALTinterwordspacing
N.~Sakimura, J.~Bradley, M.~Jones, B.~de~Medeiros, and C.~Mortimore, ``{OpenID
  Connect Core 1.0},'' OpenID Foundation, Tech. Rep., February 2014. [Online].
  Available: \url{https://openid.net/specs/openid-connect-core-1\_0-final.html}
\BIBentrySTDinterwordspacing

\bibitem{open_banking_fapi}
\BIBentryALTinterwordspacing
{OpenID Foundation}, ``{Open Banking, Open Data and Financial-grade APIs -- A
  Whitepaper for Open Banking and Open Data Ecosystem Participants Globally},''
  3 2022. [Online]. Available:
  \url{https://openid.net/wordpress-content/uploads/2022/03/OIDF-Whitepaper\_Open-Banking-Open-Data-and-Financial-Grade-APIs\_2022-03-16.pdf}
\BIBentrySTDinterwordspacing

\bibitem{fapi2}
\BIBentryALTinterwordspacing
D.~Fett, ``{FAPI 2.0 Security Profile},'' OpenID Foundation, Tech. Rep.,
  November 2022. [Online]. Available:
  \url{https://openid.net/specs/fapi-2\_0-security-02.html}
\BIBentrySTDinterwordspacing

\bibitem{fapi_formal_prove}
D.~Fett, P.~Hosseyni, and R.~Küsters, ``{An Extensive Formal Security Analysis
  of the OpenID Financial-Grade API},'' in \emph{{IEEE Symposium on Security
  and Privacy (SP)}}, 2019, pp. 453--471.

\bibitem{fapi_openbanking}
D.~Fett, ``{FAPI 2.0: A High-Security Profile for OAuth and OpenID Connect},''
  in \emph{{Open Identity Summit}}.\hskip 1em plus 0.5em minus 0.4em\relax
  {Bonn}: {Gesellschaft für Informatik e.V.}, 2021, pp. 71--82.

\bibitem{mastercard_oauth}
{Mastercard Inc.}, ``{Mastercard Open Banking US -- OAuth Connections},''
  \url{https://developer.mastercard.com/open-banking-us/documentation/financial-institution/oauth-connections/},
  accessed: 2024-09-03.

\bibitem{paypal_oauth}
{PayPal Inc.}, ``{PayPal Developer -- Authentication},''
  \url{https://developer.paypal.com/api/rest/authentication/}, accessed:
  2024-09-03.

\end{thebibliography}
%
%
%
\appendix 
\section{Appendix}\label{sec:app}
\subsection{Acronyms}
\begin{acronym}[HTTP]
    \acro{AES}{Advanced Encryption Standard}
    \acrodef{AEAD}{Authenticated Encryption with Associated Data}
    \acro{AKE}{Authenticated Key Exchange}
    \acro{AS}{Authorization Server}
    \acro{AT}{Access Token}
    \acro{BLE}{Bluetooth Low Energy}
	\acro{CA}{Certificate Authority}
	\acroplural{CA}[CAs]{Certificate Authorities}
	\acro{CCH}{Contract Clearing House}
	\acro{CP}{Charge Point}
	\acro{CPO}{Charge Point Operator}
	\acro{CPS}{Certificate Provisioning Service}
	\acro{CSR}{Certificate Signing Request}
	\acro{DAA}{Direct Anonymous Attestation}
	\acro{DH}{Diffie Hellman}
	\acro{DNS}{Domain Name System}
	\acro{DSA}{Digital Signature Algorithm}
	\acro{DSO}{Distribution System Operator}
	\acro{ECC}{Elliptic Curve Cryptography}
	\acro{ECDH}{Elliptic Curve Diffie Hellman}
	\acro{ECDSA}{Elliptic Curve Digital Signature Algorithm}
	\acro{ECU}{Electronic Control Unit}
	\acro{EIM}{External Identification Means}
	\acro{EMAID}[eMAID]{e-Mobility Account Identifier}
	\acro{EMSP}[eMSP]{e-Mobility Service Provider}
	\acro{EV}{Electric Vehicle}
	\acro{EVCC}{Electric Vehicle Communication Controller}
    \acro{FAPI}{Financial Grade API}
	\acro{HEMS}{Home Energy Management System}
	\acro{HSM}{Hardware Security Module}
	\acro{HTTP}{Hypertext Transfer Protocol}
    \acro{IdP}{Identity Provider}
	\acro{IDS}{Intrusion Detection System}
	\acro{IoT}{Internet of Things}
    \acro{JSON}{JavaScript Object Notation}
	\acrodef{KEM}{Key Encapsulation Mechanism}
	\acro{KDF}{Key Derivation Function}
    \acro{LMS}{Leighton Micali Signature}
	\acro{MadIoT}{Manipulation of demand via IoT}
	\acro{MitM}{Man-in-the-Middle}
	\acro{MV}{Medium Voltage}
    \acrodef{NIST}{National Institute of Standards and Technology}
	\acro{OEM}{Original Equipment Manufacturer}
	\acro{OCHP}{Open Clearing House Protocol}
	\acro{OCPI}{Open Charge Point Interface}
	\acro{OCPP}{Open Charge Point Protocol}
	\acro{OICP}{Open InterCharge Protocol}
	\acro{OIDC}{OpenID Connect}
	\acro{OOB}{Out-of-Band}
    \acro{OPC UA}{Open Platform Communications Unified Architecture}
    \acro{OQS}{Open Quantum Safe}
	\acro{OSCP}{Open Smart Charging Protocol}
    \acro{PAR}{Pushed Authorization Request}
	\acro{PCR}{Platform Configuration Register}
	\acro{PCID}{Provisioning Certificate Identifier}
    \acro{PKCE}{Proof Key for Code Exchange}
	\acro{PKI}{Public Key Infrastructure}
	\acro{PLC}{Power Line Communication}
	\acro{PnC}{Plug-and-Charge}
	\acro{PoP}{Proof-of-Possession}
	\acro{PQ}{Post-Quantum}
	\acro{PQC}{Post-Quantum Cryptography}
    \acro{PR}{Protected Resource}
    \acro{RAR}{Rich Authorization Request}
    \acro{RO}{Resource Owner}
    \acro{RS}{Resource Server}
	\acro{SECC}{Supply Equipment Communication Controller}
	\acro{SoC}{State of Charge}
	\acro{SSI}{Self-Sovereign Identity}
	\acro{TLS}{Transport Layer Security}
	\acro{TPM}{Trusted Platform Module}
    \acro{UA}{User Agent}
	\acro{V2G}{Vehicle to Grid}
    \acro{V2V}{Vehicle to Vehicle}
    \acro{VAS}{Value-Added Services}
    \acro{WOTS}{Winternitz One-Time Signature}
    \acro{XMSS}{eXtended Merkle Signature Scheme}
\end{acronym} 
\subsection{Extension: Single Sign-On with Payment Providers}
\label{sec:oidc_extension}
This section provides technical details on the \ac{OIDC} extension, whose usability benefits are discussed in \sect{concept_oidc}.

\begin{figure}
    \centering
    \includegraphics[width=1.0\linewidth]{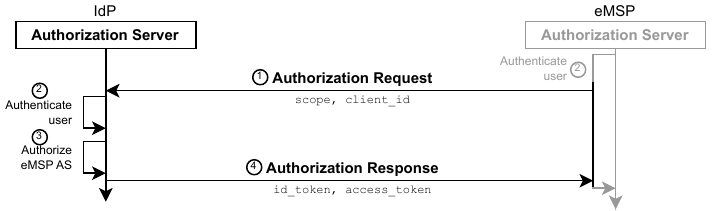}
    \caption{Simplified \acf{OIDC} authorization flow.}
    \label{fig:oidc_flow}
\end{figure}

Modern Single Sign-On (SSO) systems are commonly implemented using the \acf{OIDC} standard  \cite{OidcCore}, which is built upon OAuth 2.0 (\rfc{6749}). A simplified overview of how \ac{OIDC} can be integrated into step 2 of the authorization process (as outlined in \sect{authorization_request}) is shown in \fig{oidc_flow}.

In this scenario, instead of manually entering login credentials for an \ac{EMSP} account, the driver opts to authenticate via a third-party \ac{IdP}.
This initiates an Authorization Request to the selected \ac{IdP}’s \ac{AS} (1), redirecting the driver (i.e., the \ac{RO}) to the \ac{IdP}'s \ac{AS} through their web browser. The driver then authenticates using their existing \ac{IdP} credentials (2) and authorizes the \ac{EMSP}'s \ac{AS} to access specific profile information—such as name and address—using the scope \texttt{openid profile address}. If the \ac{IdP} is also a payment provider, the Authorization Request may include an additional scope to request continuous payment authorization. Upon granting the requested scope (3), the \ac{IdP}'s \ac{AS} issues an access token to the \ac{EMSP}'s \ac{AS} (4), allowing the \ac{EMSP} to execute payments. Simultaneously, the \ac{IdP}'s \ac{AS} provides an ID token (4) containing the driver’s profile information. Both tokens are signed by the \ac{IdP}'s \ac{AS}.

Such payment \acp{IdP} are typically private online payment providers which are often used in online shops, e.g., Apple (Pay), Google (Pay), or PayPal.
However, the OpenID Foundation proposed the use of a bank's \ac{AS} as \acp{IdP} for OAuth-based payment authorizations in its Open Banking whitepaper \cite{open_banking_fapi}.
For this, the OpenID Foundation has developed the \ac{FAPI} security profile \cite{fapi2} which was formally analyzed by Fett et al. \cite{fapi_formal_prove}.
\ac{FAPI} is already adopted by large ecosystems like OpenBanking UK \cite{fapi_openbanking}.
Additionally, payment providers like MasterCard \cite{mastercard_oauth} and PayPal \cite{paypal_oauth} have integrated OAuth 2.0 into their systems.

Certain \acp{EMSP} may already offer Single Sign-On capabilities and online payment authorization capabilities within the ISO 15118-based \ac{PnC} framework, as these technologies are complementary. However, the proposed OAuth-based extension to the ISO 15118 \ac{PnC} authorization process significantly enhances user convenience, enabling rapid switching between \ac{EMSP}. This flexibility allows drivers to take advantage of discounts at \acp{CP} offered by specific \acp{EMSP}, while simultaneously lowering the barriers for \acp{EMSP} to attract new customers by simplifying contract initiation.

\subsection{Example of an Injective Agreement Lemma in Tamarin}\label{sec:app:tam}
\cref{fig:eval-inj} shows an example of a Tamarin lemma used to prove injective agreement between the \ac{EMSP} and the \ac{EV} for credential installation. This, and other lemmas, are part of our full Tamarin model definition, which can be found online (cf. \cref{f:git} on \cpageref{f:git}).

\definecolor{mygreen}{rgb}{0,0.5,0}
\definecolor{mymauve}{rgb}{0.58,0,0.82}

\lstdefinelanguage{tamarin}
{
	alsoletter={-}, 
	morekeywords={
		equations,
		functions,
		builtins,
		protocol,
		property,
		theory,
		begin,
		end,
		subsection,
		section,
		text,
		rule,
		pb,
		lts,
		exists-trace,
		all-traces,
		enable,
		assertions,
		modulo,
		default\_rules,
		anb-proto,
		in,
		let,
		Fresh,
		fresh,
		Public,
		public,
		restriction,
		lemma
	},
	keywordstyle=\color{mygreen},
	otherkeywords={
		=,
		@,
		<,
		>,
		|,
		],
		[,
		]-, 
		--, 
		==>,
		<=>,
		--[,
		]->,
		-->,
		.,
		\&,
		",
		:,
		\^,
		\$,
		!,
		~
	},
	morekeywords=[2]{
		hashing,
		h,
		asymmetric-encryption,
		aenc,
		adec,
		pk,
		signing,
		sign,
		verify,
		true,
		revealing-signing,
		revealSign,
		revealVerify,
		getMessage,
		symmetric-encryption,
		senc,
		sdec,
		diffie-hellman,
		inv,
		\^,
		1,
		bilinear-pairing,
		pmult,
		em,
		xor,
		zero,
		multiset
	},
	keywordstyle=[2]\color{teal}\textbf,
	morekeywords=[3]{
		=,
		@,
		<,
		>
	},
	keywordstyle=[3]\color{orange},
	morekeywords=[4]{
		Ex,
		All,
		F,
		T,
		|,
		.,
		\$,
		!,
		~,
		\&,
		==>, 
		<=>
	},
	keywordstyle=[4]\color{mymauve},
	morecomment=[l]{//}, 
	morecomment=[s]{/*}{*/}, 
	commentstyle=\color{teal},
	morestring=[b]', 
	stringstyle=\color{mymauve},
	breaklines=true,
	tabsize=2,
	xleftmargin=2em,
	basicstyle={\small}
}
\lstset{language=tamarin}
\begin{figure}
\begin{lstlisting}[numbers=left, breaklines, tabsize=2, caption={Tamarin Lemma for Injective Agreement Between the \ac{EMSP} and the \ac{EV} for Credential Installation}, captionpos=b, label={fig:eval-inj}]
lemma auth_inj_agreement_EMSP_EV:
"
All emsp evId PKev #i .
CommitEMSP_EV(emsp, evId, PKev) @ i 
==>			
Ex #j . 	
	RunningEV_EMSP(evId, emsp, PKev) @ j
	& (#j<#i) 
	& not(
		  Ex emsp2 evId2 #i2 . ( 
			CommitEMSP_EV(emsp2, evId2, PKev) @ i2 & not(#i2=#i) 
	)	)
| 
Ex RevealEvent Entity #kr . KeyReveal(RevealEvent, Entity) @ kr & Honest(Entity) @ i
"
\end{lstlisting}
\end{figure}
Intuitively, the lemma can be described as:
Whenever an \ac{EMSP} accepts a public key from an \ac{EV} (line~3--4) it follows that the same \ac{EV} has previously sent an installation request with the same public key to the same \ac{EMSP} (lines~5--8). Moreover, each accepted public key by the \ac{EMSP} corresponds to a unique request (lines~9--12).
These security properties may only be violated if the long term secrets of one of the parties, involved in the specific protocol run, were leaked (line~14).
Injective agreement in the reverse direction is modeled analogously over the contract certificate generated by the \ac{EMSP} and accepted by the \ac{EV}.

\end{document}